\documentclass[pdflatex,sn-mathphys-num]{sn-jnl}

\usepackage{caption, subcaption, graphicx}%
\usepackage{multirow}%
\usepackage{amsmath,amssymb,amsfonts}%
\usepackage{amsthm}%
\usepackage{mathrsfs}%
\usepackage[title]{appendix}%
\usepackage{xcolor}%
\usepackage{textcomp}%
\usepackage{manyfoot}%
\usepackage{booktabs}%
\usepackage{algorithm}%
\usepackage{algorithmicx}%
\usepackage{algpseudocode}%
\usepackage{listings}%
\usepackage{siunitx}%

\usepackage{placeins}

\theoremstyle{thmstyleone}%
\theoremstyle{thmstyletwo}%

\theoremstyle{thmstylethree}%

\usepackage{lineno}

\begin{document}

\title[Article Title]{The superconducting diode effect in Josephson junctions fabricated from structurally chiral Mo$_3$Al$_2$C}


\author*[1,2]{\fnm{Peter T.} \sur{Orban}}\email{porban1@jhu.edu}

\author[1,2]{\fnm{Gregory} \sur{Bassen}}

\author[1,2]{\fnm{Evan N.} \sur{Crites}}

\author[1]{\fnm{Takumi} \sur{Matsuo}}

\author[2]{\fnm{Maxime A.} \sur{Siegler}}

\author*[1,2,3]{\fnm{Tyrel M.} \sur{McQueen}}\email{mcqueen@jhu.edu}

\affil[1]{\orgdiv{Institute for Quantum Matter, William H. Miller III Department of Physics and Astronomy}, \orgname{The Johns Hopkins University}, \orgaddress{\street{3400 N. Charles St.}, \city{Baltimore}, \postcode{21218}, \state{Maryland}, \country{USA}}}

\affil[2]{\orgdiv{Department of Chemistry}, \orgname{The Johns Hopkins University}, \orgaddress{\street{3400 N. Charles St.}, \city{Baltimore}, \postcode{21218}, \state{Maryland}, \country{USA}}}

\affil[3]{\orgdiv{Department of Materials Science and Engineering}, \orgname{The Johns Hopkins University}, \orgaddress{\street{3400 N. Charles St.}, \city{Baltimore}, \postcode{21218}, \state{Maryland}, \country{USA}}}


\abstract{The superconducting diode effect occurs in superconducting materials in which both spin and inversion symmetry are broken. The recently observed chirality-induced spin selectivity effect demonstrates that chiral materials break both symmetries. Thus a Josephson junction interface with the left-handed structure on one side of the junction and the right-handed structure on the other should exhibit a diode effect. Here, we report the electrical transport properties of right-handed/left-handed and right-handed/right-handed devices fabricated from single crystals of the structurally chiral superconductor Mo$_3$Al$_2$C. Fraunhofer-like magnetic diffraction patterns confirm the presence of Josephson effect in all but one of our devices. A magnetic-field-induced superconducting diode effect is demonstrated in the right-handed/left-handed devices by a statistically significant difference in $I_{c+}$ and $I_{c-}$, with a maximum asymmetry of 5\%. The intrinsic superconducting diode effect is not observed in the right-handed/right-handed devices. We provide an explanation for the presence of the superconducting diode effect in the right-handed/left-handed devices. }




\maketitle

\section*{Introduction}\label{sec1}

Nonreciprocal charge transport is the underlying phenomenon of many important electronic devices, such as diodes and transistors \cite{nagaosa2024nonreciprocal}. One of the simplest examples of such a device is a \textit{p-n} junction, which is formed by the interface of a \textit{p}-type semiconductor and a \textit{n}-type semiconductor \cite{Shockley1949}. This inversion symmetry breaking device has an asymmetric current-voltage characteristic in zero magnetic field, making this ideal for applications in logic and computation \cite{sze2021physics}. 

The superconducting analogue, known as the Josephson diode, is a device that shows single-directional superconductivity, and theorists have proposed several designs to observe this effect \cite{PhysRevB.103.245302, PhysRevLett.99.067004, PhysRevB.98.075430}. To observe the superconducting diode effect (SDE), time-reversal symmetry and inversion symmetry must be broken simultaneously \cite{tokura2018nonreciprocal}. The SDE has been observed in noncentrosymmetric superconductors \cite{ando2020observation, wakatsuki2017nonreciprocal} and 2D electron gases \cite{baumgartner2022supercurrent}, but both require an external applied magnetic field to break time-reversal symmetry. Recently, a magnetic-field-free Josephson diode was observed in NbSe$_2$/Nb$_3$Br$_8$/NbSe$_2$ devices \cite{wu2022field}, and theoretical work suggests that this effect occurs due to an electric polarization in the Nb$_3$Br$_8$ layer inducing spin selectivity (time-reversal symmetry breaking) when there is electrical transport \cite{zhang2022general}. A high temperature Josephson diode has also been observed in a device with twisted Bi$_2$Sr$_2$CaCu$_2$O$_8$ (BSCCO) flakes with a small applied magnetic field \cite{ghosh2024high}.

The chirality-induced spin selectivity (CISS) effect, discovered in 1999, is the spin-selective electron scattering by chiral (i.e. a specific subclass of noncentrosymmetric materials with only rotational symmetries) molecules, even without unpaired spins or externally applied magnetic fields (two standard time-reversal symmetry breaking mechanisms) \cite{bloom2024chiral, naaman2012chiral}. We thus wondered whether a magnetic-field-free SDE can be generated by forming junctions between materials of opposite handedness. In addition, the superconducting diode effect has been proposed to exist in Josephson junctions with a chiral quantum dot \cite{cheng2023josephson, debnath2024gate, debnath2025field} or chiral nanotube \cite{he2023supercurrent} as the barrier, making chirality a viable platform for observing the superconducting diode effect. Here, we experimentally explore using a superconductor with a chiral crystal structure to fabricate a Josephson junction where one side of the junction is left-handed and the other side is right-handed. This idea is illustrated in Figure 1a. Unfortunately, there are no known 2D superconductors with a chiral crystal structure, so it is not possible to use conventional techniques such as exfoliation and the polydimethylsiloxane assistant dry transfer method to fabricate this type of junction \cite{castellanos2014deterministic}. Thin film deposition techniques are also challenging, since it is extremely difficult to control atomic-scale handedness of materials in thin film form \cite{kim2016chiral}. 

Instead, we adopt techniques used to make cuprate and iron pnictide Josephson junctions, which involves applying pressure to push two bulk crystals together \cite{zhang2009josephson, zhang2009observation}. This technique allows for fabrication of Josephson junctions with relatively symmetric magnetic diffraction patterns out of bulk 3D materials. However, there is less control over the geometry of the junction, since the intended overlap areas and effective junction areas differ due to more oxidized areas on the crystal surface that do not allow superconducting tunneling \cite{zhang2009observation, zhang2009josephson}. Therefore, only a small fraction of the contact areas are responsible for the Josephson effect, and this junction size can be estimated from the period of the critical current modulation as a function of magnetic field.

The structurally chiral superconductor Mo$_3$Al$_2$C crystallizes in the cubic space groups P4$_1$32 and P4$_3$32 (Figure 1b) and has a transition temperature of $T_{c} = 9.2$~K, making it possible to study this system at $^4$He temperatures \cite{bauer2010unconventional, zhigadlo2018crystal}. Although this superconductor was originally thought to be unconventional, it has been shown to be a strongly coupled s-wave superconductor \cite{bonalde2011evidence}. Recent Raman studies show that Mo$_3$Al$_2$C undergoes a structural transition at $T = 155$~K to a rhombohedral structure making the crystal structure polar and chiral \cite{wu2024polar}. However, for simplicity, we will label these crystals by their room temperature structure as observed by single crystal X-ray diffraction. In agreement with previous work on the structurally chiral superconductor NbGe$_2$ \cite{li2025absence}, we find no SDE in individual single-handedness crystals of Mo$_3$Al$_2$C. We fabricate a Josephson junction by pressing a P4$_1$32 (right-handed) and a P4$_3$32 (left-handed) crystal of Mo$_3$Al$_2$C together with a 3D-printed pressure device. We show that a field-induced diode effect can be observed in the right-handed/left-handed devices, but there is no intrinsic diode effect in the right-handed/right-handed devices. Fraunhofer-like patterns confirm the presence of the Josephson effect in all but one of our devices. The current-voltage characteristic for the right-handed/left-handed devices shows a critical current where $I_{c+} \neq I_{c-}$ at finite magnetic fields. Multiple critical current measurements show that this effect is robust, and that the direction of the diode effect switches as a function of magnetic field. Future work is required to determine if a zero field diode effect occurs in right-handed/left-handed devices, and we encourage future work on Josephson junctions fabricated from structurally chiral superconductors.

\section*{Results}\label{sec2}
In Figure 1c, a scanning electron microscope (SEM) image of a single crystal of Mo$_3$Al$_2$C shows that these crystals have flat surfaces, making it possible to fabricate devices directly from bulk crystals. The electrical transport properties of a single crystal of Mo$_3$Al$_2$C (device 1) are summarized in Figure 2. A drop in the resistance is observed at the known transition temperature of $T_{c} = 9$~K \cite{zhigadlo2018crystal}, and a second transition is observed at $T_{c2} = 3$~K due to inclusions of Mo$_2$C. The current-voltage relation at $T = 1.8$~K shows a critical current, which we define to be the current where $dV/dI$ is maximized, of $I_{c+} = 2.48$~mA and $I_{c-} = -2.48$~mA, which is consistent with previous work \cite{zhigadlo2018crystal}. We note that due to the two probe nature of the device, there is a finite resistance in the superconducting state, which is also shown by a finite slope in the superconducting region of the IV curve. Magnetic susceptibility and electrical resistance measurements confirm that these crystals are superconducting (see SI Figure 2a and SI Figure 3). We attribute the nonzero resistance to the contact resistance from our two probe setup, since the current/voltage terminals are touching on each side of the junction prior to intersecting with the superconducting crystal \cite{dykaar1987picosecond}. 

We fabricated Josephson junctions by pressing together the (110) planes of single crystals of Mo$_3$Al$_2$C to make right-handed/left-handed and right-handed/right-handed devices. The (110) plane is used, since this is the naturally exposed crystallographic plane for plate-like crystals of Mo$_3$Al$_2$C \cite{wu2024polar}. The device making process is shown in Figures 1c-1e and is discussed in detail in the methods section. The insulating layer of the device is the nominal oxide layer on the crystal, which is of order 10~\AA~\cite{evertsson2015thickness}. At zero applied magnetic field, we measured the resistance of two right-handed/left-handed (device 2 and device 3) and two right-handed/right-handed (device 4 and device 5) devices down to base temperature, $T = 1.8$~K (Figure 3). All devices show an initial decrease in resistance around $T_{c} = 9$~K, followed by a more significant drop at 6~K. The lower transition temperature in the device compared to the bulk $T_{c}$ is expected behavior \cite{tian2021josephson} due to the presence of the insulating barrier. 

At $T = 1.8$~K, we investigate the voltage versus current characteristic of the devices at zero applied magnetic field. We note that this is not a true zero-field measurement, as it is possible that there is a remnant field in our Physical Properties Measurement System. The current is swept in four branches: 0 to $I_{\text{max}}$,  $I_{\text{max}}$ to 0, 0 to $-I_{\text{max}}$, and $-I_{\text{max}}$ to 0, and the IV characteristics for the right-handed/left-handed devices and right-handed/right-handed devices are shown in Figure 4a-4d. The same curve is obtained when the current is swept in the reverse direction (see SI Figure 6). Due to contact resistance, the superconducting state has a non-zero voltage, but as shown by plotting $dV/dI$ versus $I$, the critical current is still evident. The critical currents are summarized in Table 1. These critical currents are significantly smaller than the critical current of a single crystal of Mo$_3$Al$_2$C, as expected for two weakly linked superconductors. We have observed this effect in other devices (device 6, device 7, and device 8) of both man-made and naturally formed twin junctions (SI Figures 14-22). 

Although these devices have contact resistance, we can estimate the $I_{c} R_{n}$ product. We computed the slopes of the IV curves above and below $I_{c+}$ and found the difference between the two to determine $R_{n}$. The normal state resistance $R_{n}$ and the $I_{c}R_{n}$ products are summarized in Table 1. Given the superconducting gap of $2$ meV for Mo$_3$Al$_2$C \cite{bauer2014absence} and the Ambegaoakar-Baratoff relation for an ideal conventional superconductor-insulator-conventional superconductor junction, the predicted $I_{c}R_{n}$ for an ideal Mo$_3$Al$_2$C junction is $I_{c}R_{n} \approx \frac{\pi \Delta}{2e} \approx 3$ mV. The large $I_{c}R_{n}$ product can therefore be attributed to the large superconducting gap of Mo$_3$Al$_2$C.

To confirm the presence of Josephson tunneling, we measured the critical current as a function of an in-plane magnetic field. For an ideal Josephson junction in an in-plane applied magnetic field, the critical current should take the form of a Fraunhofer diffraction pattern, given by
\begin{equation}
    I_{c} = I_{0} \frac{\sin \bigg(\frac{\pi \Phi}{\Phi_0}\bigg)}{\frac{\pi \Phi}{\Phi_0}}
\end{equation}
where $I_{0}$ is the maximum critical current, $\Phi$ is the total magnetic flux experienced by the junction, and $\Phi_0$ is the magnetic flux quantum. Critical current versus applied magnetic field is shown in Figure 5 for devices 2 (right-handed/left-handed), 3 (right-handed/left-handed), 4 (right-handed/right-handed), and 5 (right-handed/right-handed). An ideal Fraunhofer pattern is not observed, yet we still see oscillations in $I_{c}$ versus $H$ that resemble Fraunhofer-like maxima in all devices except device 5 (right-handed/right-handed).

We can estimate the effective width $W$ of the junction based on the modulations of the critical current as a function of magnetic field. \cite{takeuchi1996anisotropic}. The total magnetic flux seen by the junction is given by
\begin{equation}
    \Phi \approx HW(2 \lambda_L + d_{b}),
\end{equation}
where $H$ is the magnetic field, $\lambda_L$ is the London penetration depth, and $d_b$ is the tunnel barrier thickness. Taking the reported value of $\lambda_L$ to be about 400 nm \cite{bauer2014absence} and determining the modulation period of the field $H_0$, the width of the junction can be estimated as
\begin{equation}
    W = \frac{2 \Phi_0}{H_{0}(2 \lambda_L + d_b)}.
\end{equation}
The tunnel barrier thickness $d_b$ is approximated as $1$ nm. Since $d_b << \lambda_L,$ $W$ can be estimated without precise knowledge of the barrier thickness. The estimated width for each junction is summarized in Table 2, where most of our devices have a width of order $1$ $\mu$m. Therefore, the critical current density of each device can be estimated as $J_{c} = I_{c}/W^{2}$, and these values are also summarized in Table 2. We were not able to estimate the width or current density for device 5 due to the absence of field modulations in $I_{c}$ versus $H$.

The Josephson penetration length is a length scale that describes the uniformity of supercurrent across the junction. When $W << \lambda_J$ or $W \sim \lambda_J$, the supercurrent across the junction is relatively uniform, and this is evident by fairly symmetric magnetic diffraction pattern \cite{zhang2009josephson, takeuchi1996anisotropic}. When $W >> \lambda_J$, the current supercurrent across the junction is highly nonuniform, and this results in the self-field effect. We believe that device 5 (right-handed/right-handed) is in the $W >> \lambda_J$ limit because no oscillations of $I_c$ versus $H$ are observed \cite{schwidtal1969barrier}, despite having a larger critical current, and the $I_{c}$ versus $H$ shape depends on the direction the field is swept in, as shown in Fig SI 12. The Josephson penetration length is given by
\begin{equation}
    \lambda_J = \sqrt{\frac{\Phi_0}{2 \pi \mu_0 J_c (2 \lambda_L + d_b)}},
\end{equation}
and these values are summarized in Table 2. Thus $W/\lambda_J < 1$ for devices 2, 3, and 4, suggesting that these devices are in the short to intermediate junction limit. However, no genuine Fraunhofer pattern is observed in these junctions. This could be because the magnetic diffraction pattern can depend on the shape of the area of overlap \cite{watanabe2008shape} and local fluctuations in supercurrent within the junction area \cite{zhang2009observation}. We also note that the critical current versus magnetic field for device 2 (right-handed/left-handed) is asymmetric about the central peak, and the peaks appear to be skewed, reminiscent of the self-field effect, even though $W/\lambda_J \sim 1/2$ for this device \cite{schwidtal1969barrier}. The estimated critical current density is also an order of magnitude less than devices 3 and 4, so it is likely that there is a nonuniform supercurrent in this device given its larger estimated junction size. In contrast, device 3 (right-handed/left-handed) and device 4 (right-handed/right-handed), which are closer to the short junction limit, have more symmetric peaks for the critical current as a function of magnetic field for both $I_{c+}$ and $I_{c-}$.

To explore the nonreciprocal properties of the devices, we define the asymmetry parameter $\alpha$ as
\begin{equation}
    \alpha = \frac{I_{c+} - |I_{c-}|}{I_{c+} + |I_{c-}|} \times 100 \%,
\end{equation}
and we plot $I_{c \pm}$ and $\alpha$ versus $H$ in Figure 6. In the right-handed/left-handed devices (devices 2 and 3), we notice that there is a significant difference in the shape of $I_{c+}$ versus $H$ and $I_{c-}$ versus $H$. In particular, the maximum $I_{c+}$ and $I_{c-}$ do not occur at the same field, and $I_{c+}$ versus $H$ and $I_{c-}$ versus $H$ are horizontally shifted relative to each other, resulting in a diode effect. The asymmetry shows oscillatory behavior with a maximum amplitude of about $\alpha = $ 5\%, indicating that there is a diode effect, and the direction of the diode effect switches as a function of magnetic field.

For the control device (device 4), $I_{c+}$ and $I_{c-}$ are symmetric as a function of magnetic field. Therefore, asymmetry versus magnetic field shows the absence of a diode effect. While individual points on $I_{c}$ versus $H$ show a nonzero value of $\alpha$, there is no clear trend of $\alpha$ versus $H$, which is shown for the right-handed/left-handed devices, and we attribute this to noise in the $I_{c}$ versus $H$ measurements. Device 3 (right-handed/left-handed) and device 4 (right-handed/right-handed) have similar critical current densities and junction sizes. However, only device 3 (right-handed/left-handed) shows asymmetry in $I_{c+}$ and $I_{c-}$. Therefore, we believe that the asymmetry observed in device 3 is an intrinsic superconducting diode effect rather than the self-field effect. This can be compared to the diode-like behavior in the wide junction limit devices, such as device 5 (right-handed/right-handed). There is a maximum asymmetry of approximately 5 \%, and the direction of this asymmetry switches as a function of magnetic field. The asymmetry is non-oscillatory with magnetic field and qualitatively different from that found in the right-handed/left-handed devices. We therefore attribute this asymmetry in $I_{c+}$ and $I_{c-}$ to the self-field effect.

The maximum value of the critical current can shift to nonzero field values when vortices get pinned \cite{hou2023ubiquitous}. To attempt to avoid trapping vortices in the Mo$_3$Al$_2$C electrodes, we cooled the device in zero applied magnetic field, and stayed below $H_{c1}$ when sweeping the magnetic field. We cannot rule out introducing vortices into Mo$_3$Al$_2$C since the excitation current reduces the value of $H_{c1}$. However, we observe a symmetric $I_{c+}$ and $I_{c-}$ versus $H$ for the right-handed/right-handed device (device 4) but not for the right-handed/left-handed devices (device 2 and 3). This suggests that the asymmetry does not occur due to a vortex diode effect.

To verify that this diode effect is robust, we measured IV curves for two hours at fixed fields for device 2 ($H=1$~Oe, $H=3$~Oe, and $H=6$~Oe) and device 3 ($H=-90$~Oe, $H=-30$~Oe, $H=0$~Oe, $H=10$~Oe, and $H=70$~Oe) and plotted the distribution of $I_{c+}$ and $|I_{c-}|$ in Figure 7. For device 2, with an applied field of 1~Oe, $I_{c+} >|I_{c-}|$, and the two distributions have a statistically significant difference of $22.8 \sigma$. The distributions become significantly closer at 3~Oe ($5.2 \sigma$), and at 6~Oe, the diode direction changes with a difference of $14.4 \sigma$. Similarly, for device 3, that statistical significance is $9.1 \sigma$ at $H=-90$~Oe, $5.8 \sigma$ at $H=-30$~Oe, $2.7 \sigma$ at $H=0$~Oe, $1.7 \sigma$ at $H=10$~Oe, and $6.6 \sigma$ at $H=70$~Oe. Therefore, both right-handed/left-handed devices exhibit a diode effect, that switches direction with magnetic field, exhibited by a significant difference in the distributions of $I_{c+}$ and $|I_{c-}|$. In Figure 8, we show the $I_{c+}$ and $|I_{c-}|$ distributions for device 4 for fixed fields of $H=-40$~Oe, $H=-20$~Oe, $H=0$~Oe, $H=20$~Oe, and $H=40$~Oe. At all of the measured fields, the $I_{c+}$ and $|I_{c-}|$ distributions overlap, indicating that there is no significant diode effect in the right-handed/right-handed device.

\section*{Discussion}\label{sec12}

It is thought that both inversion and time reversal symmetry breaking are necessary to produce a superconducting diode effect \cite{tokura2018nonreciprocal}. No diode effect is observed in the single crystal device or right-handed/right-handed device (device 4), even in the presence of a magnetic field. The asymmetry observed in the device 5 (right-handed/right-handed device) is attributed to the self-field effect since we do not observe oscillations in critical current versus field and the central peak is skewed \cite{schwidtal1969barrier}. Therefore, this is not considered an intrinsic diode effect. In both cases, inversion symmetry should be broken by the structural chirality of Mo$_3$Al$_2$C, and time reversal symmetry is broken by the applied magnetic field. However, no diode effect is observed. This is consistent with recent results showing the absence of a superconducting diode effect in structurally chiral NbGe$_2$ \cite{li2025absence}.

In contrast, a clear superconducting diode effect is observed in opposite chirality junctions. Inversion symmetry breaking can arise from the structural chirality of Mo$_3$Al$_2$C, and be further enhanced by having different handednesses of the crystals on opposite sides of the junction. Further possible sources of inversion symmetry breaking include a twist in the alignment of the two crystals -- the diode effect has been observed in twisted heterostructures \cite{ghosh2024high} -- and inversion symmetry breaking could arise due to a nonuniform oxide layer on the surfaces of the two crystals. Since the diode effect is only observed in the right-handed/left-handed devices but not the right-handed/right-handed devices or single chirality crystals, we postulate that the inversion symmetry breaking due to opposite chirality crystals on either side of the junction is essential for observation of the diode effect in these materials.

Further, in opposite chirality devices, at low fields, an applied external magnetic field initially enhances the magnitude of the diode effect. Future measurements are required to determine if the diode effect persists at zero-field, or whether a small external field (whether from Earth's own magnetic field, or the self-induced field from current flow) is responsible. One microscopic picture for the origin of the CISS effect breaking time reversal symmetry is that current flow induces an orbital moment to electrons, which then couples to spin via spin orbit coupling \cite{bloom2024chiral}. Our results are consistent with this picture, but we cannot rule out other origins of time reversal symmetry breaking. 

An asymmetry of $\alpha = $ 5\% is lower than the largest of superconducting diode effects seen in other devices \cite{ghosh2024high}. These experiments are performed on single crystals of Mo$_3$Al$_2$C with the (110) plane exposed, as determined from single crystal X-ray diffraction. The screw axis of Mo$_3$Al$_2$C is perpendicular to (100), (010), and (001), so it is possible that the crystal needs to have (100) exposed to maximize the superconducting diode effect in this system, i.e., have the screw axis normal to the interface of the junction. Having (100) oriented junctions may also be a more direct test of the CISS effect, since the tunneling of the Cooper pairs would be parallel to the screw axis direction. However, due to the size and shape of these single chirality crystals, it is challenging to polish the samples to expose the (100) plane, and we leave this as a potential future study.

\section*{Conclusion}\label{sec13}

We report the transport properties of superconducting weak-link devices fabricated from two opposite chirality Mo$_3$Al$_2$C single crystals and two P4$_1$32 single crystals. There is evidence for the Josephson effect shown by a Fraunhofer-like magnetic diffraction pattern for devices 2, 3, and 4, but not device 5. A magnetic field-induced diode effect is observed in the right-handed/left-handed devices, but not the right-handed/right-handed devices or single crystal device. Future experimental and theoretical work should be done on (100) oriented junctions to investigate if there is a larger diode effect in this system, and if this could possibly originate from a CISS-type effect. Our work motivates studies on Josephson junctions fabricated from structurally chiral materials, and demonstrates that Josephson junctions can be made from pressing bulk, chiral crystals together, allowing device fabrication from 3D crystal structures.

\section*{Methods}\label{sec11}

\bmhead{Synthesis}
Single crystals of Mo$_3$Al$_2$C were grown by a high-temperature flux reaction with excess Al as a flux \cite{wu2024polar}. An open alumina tube (0.25" OD, 0.156" ID) was overheated in a xenon floating zone to melt one end of the tube until the ring shape converged into a droplet of viscous alumina and then slowly cooled to room temperature over three hours to avoid cracking. Powders of Mo (Strem, Lot \#35229900, 99.95\%) and C (Sigma Aldrich, Lot \#484164-106, 99.95\%) were ground together with a mortar and pestle, and then added to the tube with Al pieces (Kurt J. Lesker, Lot \#ALFP3352113022-1, 99.99\%) in a 3:4:1 (Mo:Al:C) ratio. The tube was placed back in the xenon floating zone and sealed using the same technique under one bar of flowing argon. The sealed tube was then heated at \SI{100}{\celsius} per hour to \SI{1650}{\celsius}, held at \SI{1650}{\celsius} for 24 hours, and then slowly cooled to \SI{1250}{\celsius} at \SI{5}{\celsius} per hour before cooling to room temperature at \SI{100}{\celsius} per hour.

\bmhead{X-ray Diffraction}
A Bruker D8 Focus spectrometer was used to collect intensity versus $2 \theta$ data on ground single crystals of Mo$_3$Al$_2$C (SI Figure 1). Single crystals could be removed by mechanically breaking larger chunks of product and picking out pieces with regular shapes. For the devices, we isolated plate-like crystals, which have the (110) plane exposed. To determine the handedness of each crystal, single crystal X-ray diffraction (scXRD) data were collected at $T = 213(2)$~K using a SuperNova diffractometer (equipped with an Atlas detector) with Mo K$\alpha$ radiation with the program CrysAlisPro (Version CrystAisPro 1.71.42.49, Rigaku OD, 2022). The same program was used to refine the unit cell dimensions and for data reduction, and the crystal structures were solved using direct methods and refined using SHELXL \cite{sheldrick2015crystal} and WINGX \cite{farrugia1999wingx}.

Due to the anomalous dispersion effect, it is possible to distinguish between left-handed and right-handed structures with scXRD \cite{cianci2005anomalous}. a chiral crystal structure, the intensity $I$ for a plane $(hkl)$ is given by
\begin{equation}
    I(hkl) = (1-x)\big|F(hkl) \big|^{2} + x\big|F(-h-k-l) \big|^{2},
\end{equation}
where $x$ is the Flack parameter and $F$ is the structure factor amplitude \cite{flack1983enantiomorph}. The Flack parameter ranges from 0 to 1, where a value close to zero implies the absolute structure is correct, and a value close to 1 means the inverted structure is the correct space group. A value close to 0.5 means there is a mixture of the enantiomers. We are able to isolate single chirality crystals of approximately 100-200 microns.

\bmhead{Scanning Electron Microscopy} Scanning electron microscopy (SEM) and energy dispersive spectroscopy (EDS) were performed on single crystals of Mo$_3$Al$_2$C mounted on carbon tape with a JEOL JSM-IT100.

\bmhead{Physical Properties}
A Quantum Design MPMS3 system was used to confirm superconductivity in the Mo$_3$Al$_2$C single crystals. Magnetization versus temperature was measured from $T = $ 2-20~K with an applied field of $H = 30$~Oe. A Quantum Design PPMS with the resistivity option was used to measure resistivity versus temperature from $T = $ 2-300~K on a polycrystalline sample of Mo$_3$Al$_2$C to also confirm superconductivity.

\bmhead{Device Fabrication}
A pressure device was designed in Blender, sliced in UltiMaker Cura,  and then 3D printed with a Snapmaker J1s 3D printer using ABS filament with a layer height of 0.28~mm and an infill density of 100\%. A single crystal of Mo$_3$Al$_2$C was placed on a mica substrate, and the immersion oil from the scXRD was washed off using a drop of dichloromethane. Due to the small size of the single crystals, we adopted a pseudo 4-probe technique to make electrical contact to the device. Conventional lithography techniques are challenging due to the 3D nature of the crystals. A “v” shape of DuPont 4922N-100 silver paint was then hand-painted on to the substrate, and an Axis Pro Micromanipulator was used to position the crystal in the silver paint (Figure 1d). Instead of silver paint, some of the devices were fabricated by depositing gold contacts on the mica substrate. Gold pads were fabricated on mica using magnetron sputtering and standard liftoff techniques. A small amount of silver paint was used to connect the crystal to the gold pattern. This process was repeated for a second crystal, and then the two crystals were stacked together on the base of the pressure device (Figure 1e). The crossbar for the pressure device is screwed down to push the two crystals together, and additional wires were attached to connect the sample to a Quantum Design universal sample mount (Figure 1f). One side of the junction has V+ and I+ terminals attached, where the other has V- and I- attached.

Devices were made with one P4$_1$32 and one P4$_3$32 crystal, as well as devices with two P4$_1$32 crystals as control devices. To make sure the critical currents observed were due to the interface effects of the two crystals, we also measured a single crystal device as a comparison. The single crystal device was fabricated using the same methods described above, where the top substrate is a blank "v" of silver paint which is pressed onto the bottom crystal.

\bmhead{Transport Measurements}
A Quantum Design PPMS with the horizontal rotator probe and AC transport option were used to investigate the transport properties of these devices. AC resistance was measured from $T =300$~K to $T = 1.8$~K with an excitation current of 0.01~mA and frequency of 17~Hz to confirm the superconducting transition of the device. At base temperature, IV curves were measured to determine the critical current of the device. The critical current as a function of an in-plane magnetic field was determined for fields $-20 \leq H \leq 20$~Oe (right-handed/left-handed device) or $-100 \leq H \leq 100$~Oe (right-handed/right-handed) to show evidence for the Josephson effect. IV curves were measured at fixed magnetic fields for two hours at each magnetic field to show the robustness of the diode effect. IV curves were also measured at temperatures from $10$~K down to $1.8$~K to show that the observed critical current is related to the superconductivity of the devices.

\section*{Data availability}
The data from this work are available at (insert DOI here).
\section*{Acknowledgements}
This work was funded by the U.S. Department of Energy, Office of Science, National Quantum Information Science Research Centers, Co-Design Center for Quantum Advantage (C2QA) under contract number DE-SC0012704. The MPMS3 system used for magnetic characterization was funded by the National Science Foundation, Division of Materials Research, Major Research Instrumentation Program, under Award No, 1828490. G.B. acknowledges support from the Harry and Cleio Greer Fellowship.

The authors would like to thank Eli Zoghlin for help with initial efforts to seal alumina tubes in the xenon floating zone.

\section*{Author Contributions}
T.M.M and P.T.O conceived of the experiment. P.T.O synthesized the materials. P.T.O and M. A. S. collected single crystal XRD data to sort left-handed and right-handed crystals. G.B. collected and analyzed the SEM data. E.N.C. and P.T.O. designed and 3D printed the pressure devices. T.M. prepared the gold leads on the mica substrates. P.T.O fabricated and performed transport measurements on the devices. P.T.O. and T.M.M analyzed the data, interpreted the results, and wrote the manuscript.

\section*{Competing Interests}
The authors declare no competing interests.

\begin{appendices}




\end{appendices}


\bibliography{sn-bibliography}

\newpage
\section*{Figures}\label{sec6}
\newpage

\begin{figure}
\centering
\includegraphics[width= \linewidth]{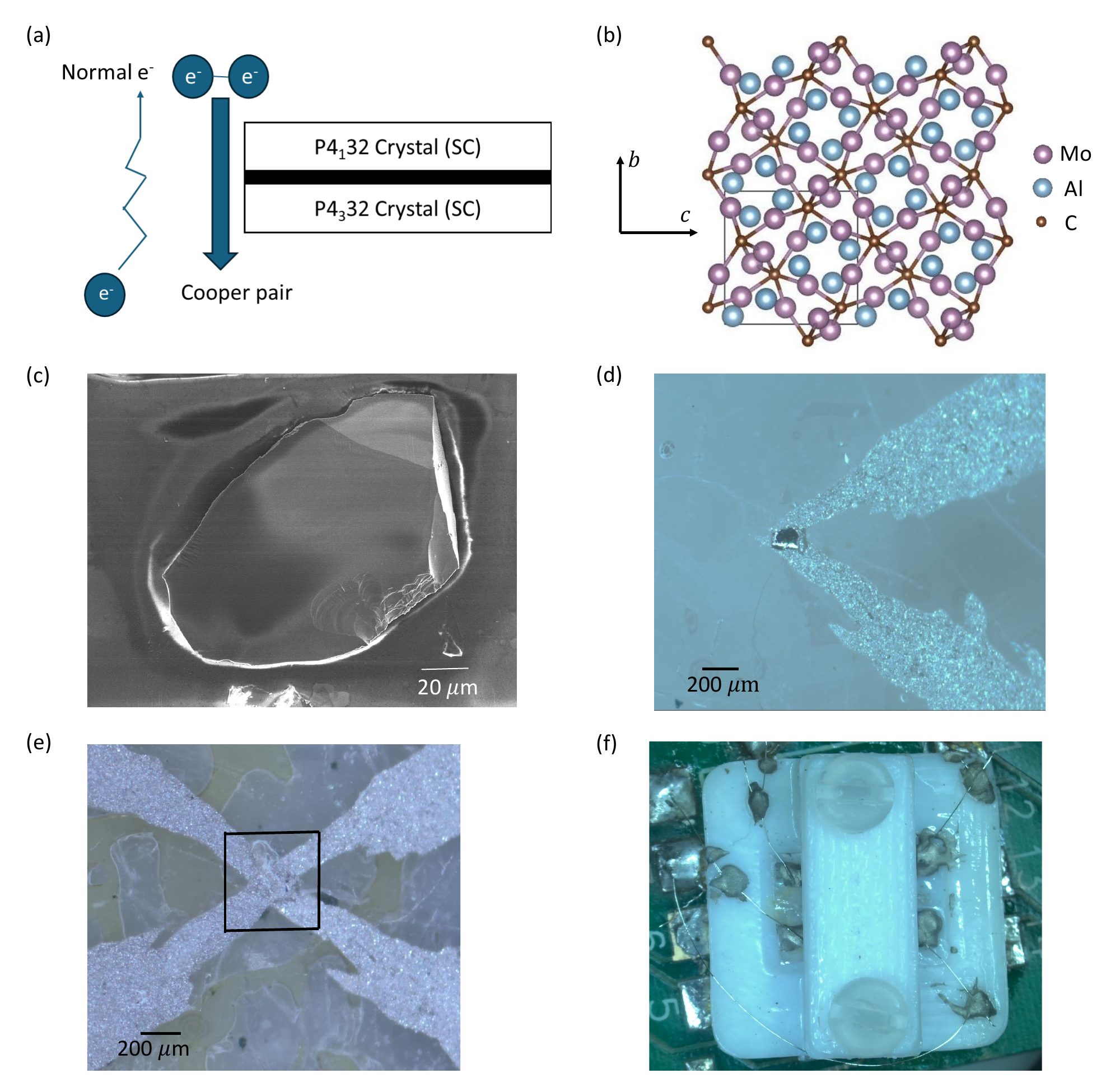}
\caption{(a) Crystal structure of Mo$_3$Al$_2$C looking along the $a$ axis. (b) Schematic diagram of a vertical Josephson junction with a right-handed (P4$_1$32) and left-handed (P4$_3$32) crystal. (c) An SEM image of a single crystal of Mo$_3$Al$_2$C. (d) A single crystal of Mo$_3$Al$_2$C positioned on a "v" of silver paint on a mica substrate to make electrical contact. (e) Two Mo$_3$Al$_2$C crystals stacked together in preparation to be pressed together. The crystals are located where the silver paint for each crystal overlaps, shown by the black box. (f) The final device with the crossbar screwed down to push the two crystals together and with electrical connections made to a Quantum Design PPMS universal sample mount. }
\label{fig:enter-label}
\end{figure}

\begin{figure}
\centering
\includegraphics[width= 0.9 \linewidth]{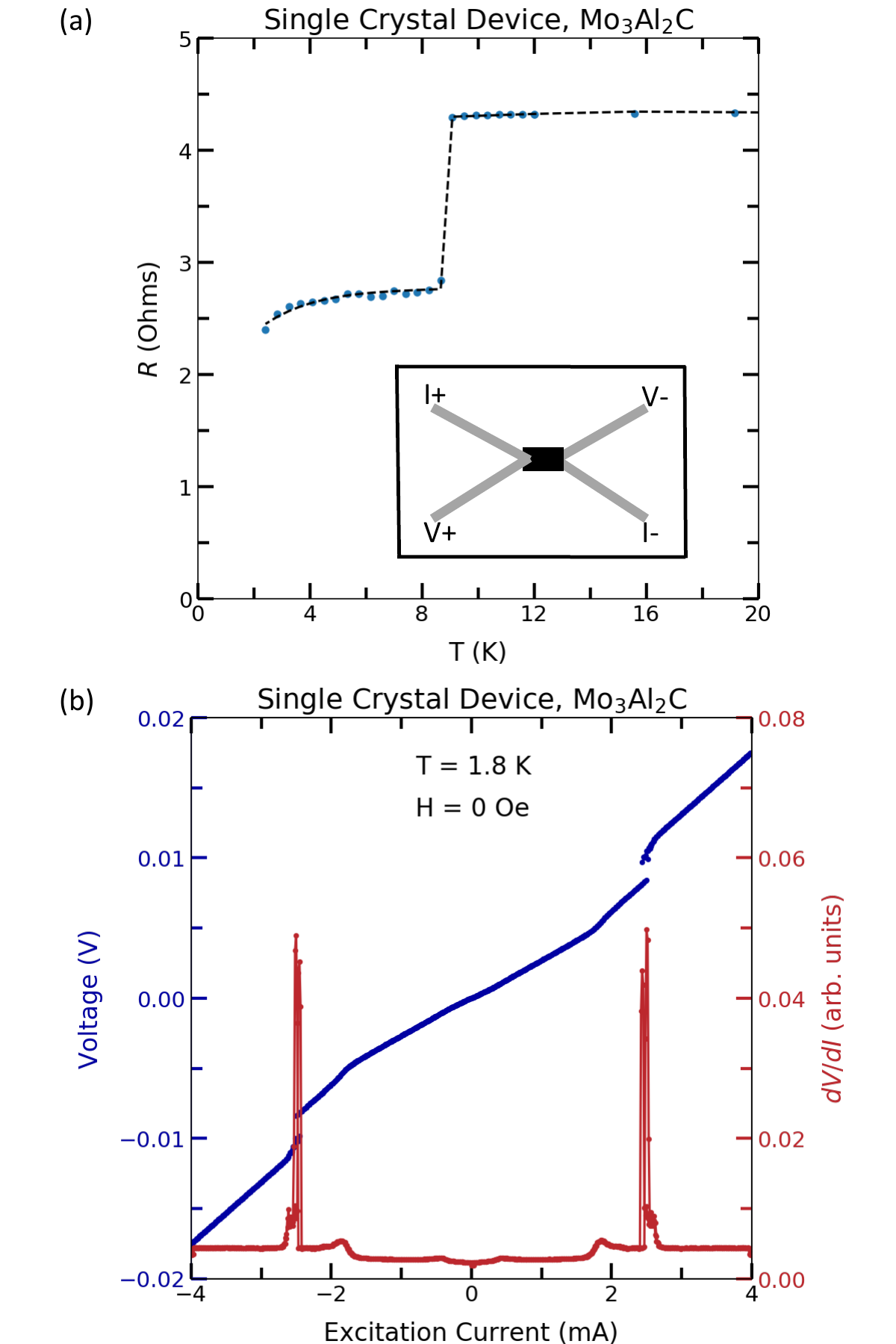}
\caption{(a) Resistance versus temperature and (b) voltage and $dV/dI$ versus excitation current for a single crystal device. The inset in (a) shows the electrical connections made to the device, where there is a current and voltage lead on each side of the junction.}
\label{fig:enter-label}
\end{figure}

\begin{figure}
\centering
\includegraphics[width=\linewidth]{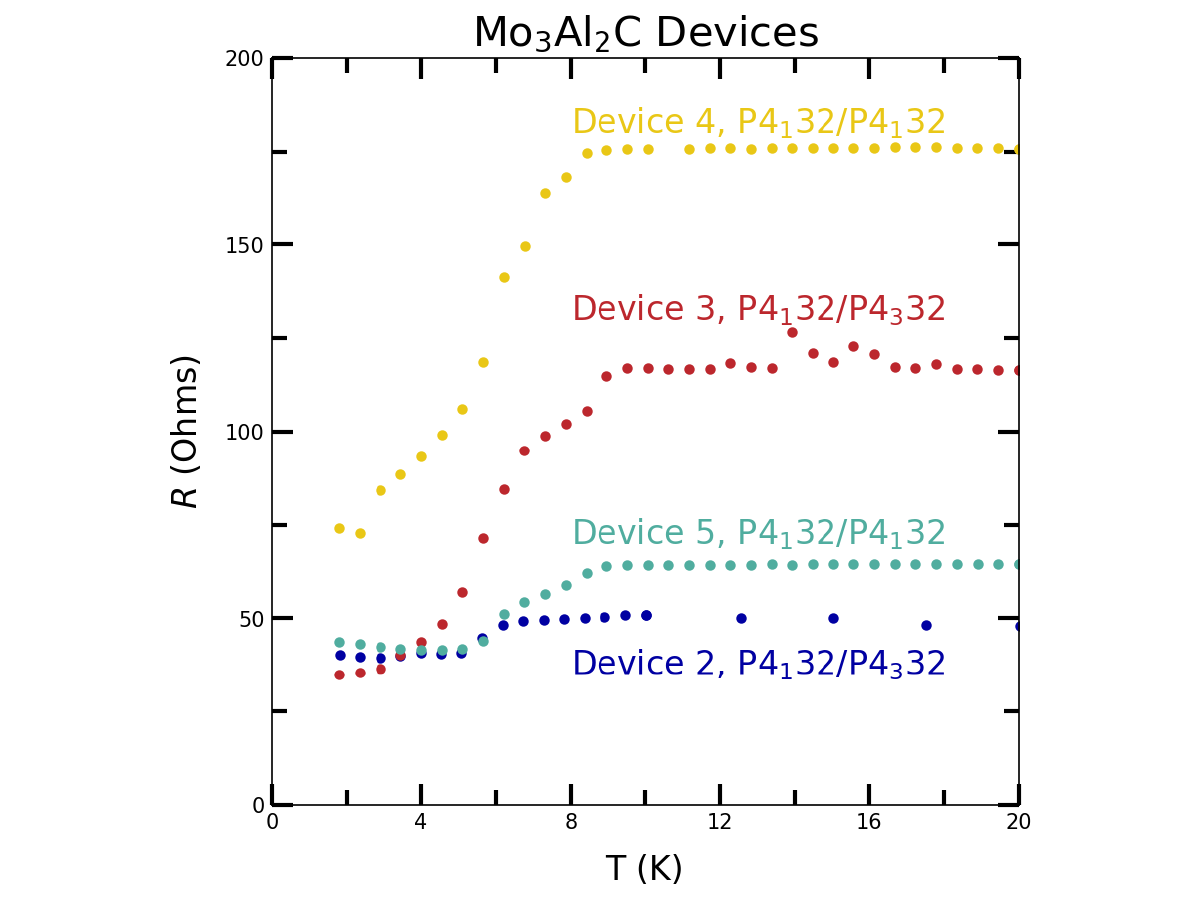}
\caption{Resistance versus temperature for the P4$_1$32/P4$_3$32 and P4$_1$32/P4$_1$32 devices.}
\label{fig:enter-label}
\end{figure}

\begin{figure}
\centering
\includegraphics[width= 1.1 \linewidth]{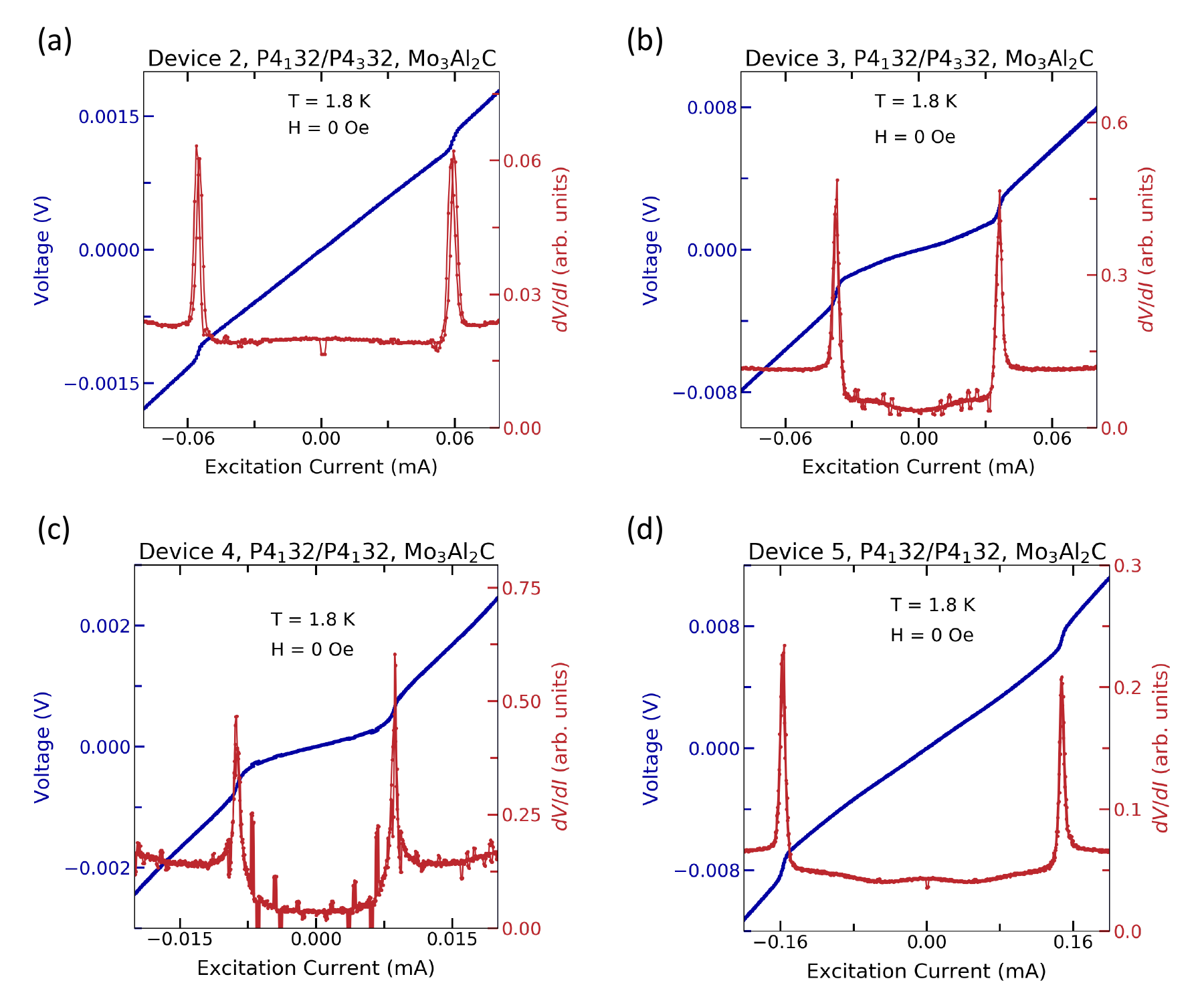}
\caption{Current-voltage relation and $dV/dI$ versus excitation current for (a) Device 2 (P4$_1$32/P4$_3$32), (b) Device 3 (P4$_1$32/P4$_3$32), (c) Device 4 (P4$_1$32/P4$_1$32), and (d) Device 5 (P4$_1$32/P4$_1$32). The current is swept in four branches: $0$ to $I_{\text{max}}$, $I_{\text{max}}$ to $0$, $0$ to $-I_{\text{max}}$, and $-I_{\text{max}}$ to $0$.}
\label{fig:enter-label}
\end{figure}

\FloatBarrier
\begin{table}[h]
\caption{A summary of the critical current in the positive ($I_{c+})$ and negative ($I_{c-}$) directions, normal state resistance $R_{n}$, and $I_{c}R_{n}$ product for each device. }\label{tab1}%
\begin{tabular}{@{}lllll@{}}
\toprule
Device Number & $I_{c+}$ (mA) & $I_{c-}$ (mA) & $R_{n}$ ($\Omega$) & $I_{c}R_{n}$ (mV) \\
\midrule
2    &  0.058  & -0.054  & 4.6 & 0.26  \\
3   &  0.036  & -0.036 &  74.8 & 2.69  \\
4   &  0.0081  & -0.0087 &  108.9 & 0.88  \\
5    &  0.147  & -0.155 & 21.9 & 3.21  \\\botrule
\end{tabular}

\end{table}
\FloatBarrier

\begin{figure}
\centering
\includegraphics[width= 1.1 \linewidth]{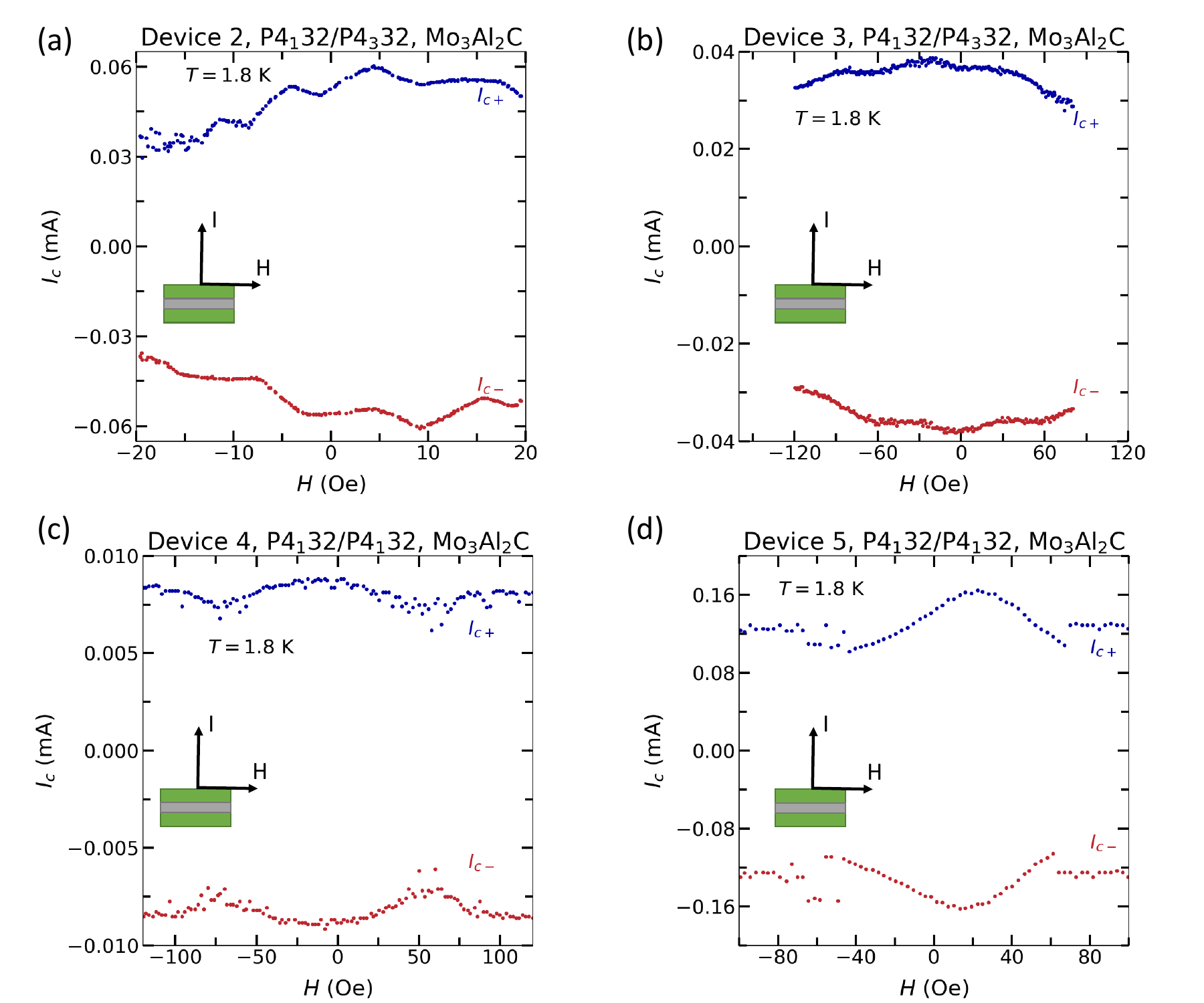}
\caption{Critical current as a function of an in-plane applied magnetic field at $T = 1.8$~K for (a) Device 2 (P4$_1$32/P4$_3$32), (b) Device 3 (P4$_1$32/P4$_1$32), (c) Device 4 (P4$_1$32/P4$_1$32), and (d) Device 5 (P4$_1$32/P4$_1$32). The inset shows the geometry of the measurement where the green and gray layers represent the superconducting and insulating layers, respectively.}
\label{fig:enter-label}
\end{figure}

\FloatBarrier
\begin{table}[h]
\caption{A summary of relevant junction parameters for each device. $H_{0}$ is the approximate period of oscillation in $I_{c}$ versus $H$, $W$ is the estimated junction width, $J_{c}$ is the estimated critical current density, and $\lambda_J$ is the estimated Josephson penetration length. }\label{tab1}%
\begin{tabular}{@{}llllll@{}}
\toprule
Device Number & $H_{0}$ (Oe) & $W$ ($\mu$m) & $J_{c}$ (A/m$^{2}$) & $\lambda_J$ ($\mu$m) & $W/\lambda_J$ \\
\midrule
2    &  10  & 5  & $2.4 \times 10^{6}$ & 10 & $\sim$ 1/2 \\
3   &  60  & 1 &  $3.7 \times 10^{7}$ & 3.0 & $\sim$ 1/3 \\
4    &  120  & 0.4 & $4.7 \times 10^{7}$ & 2.6 & $\sim$ 1/6 \\
5    &  --  & -- & -- & -- & $>>$ 1 \\\botrule
\end{tabular}

\end{table}
\FloatBarrier

\begin{figure}
\centering
\includegraphics[width=\linewidth]{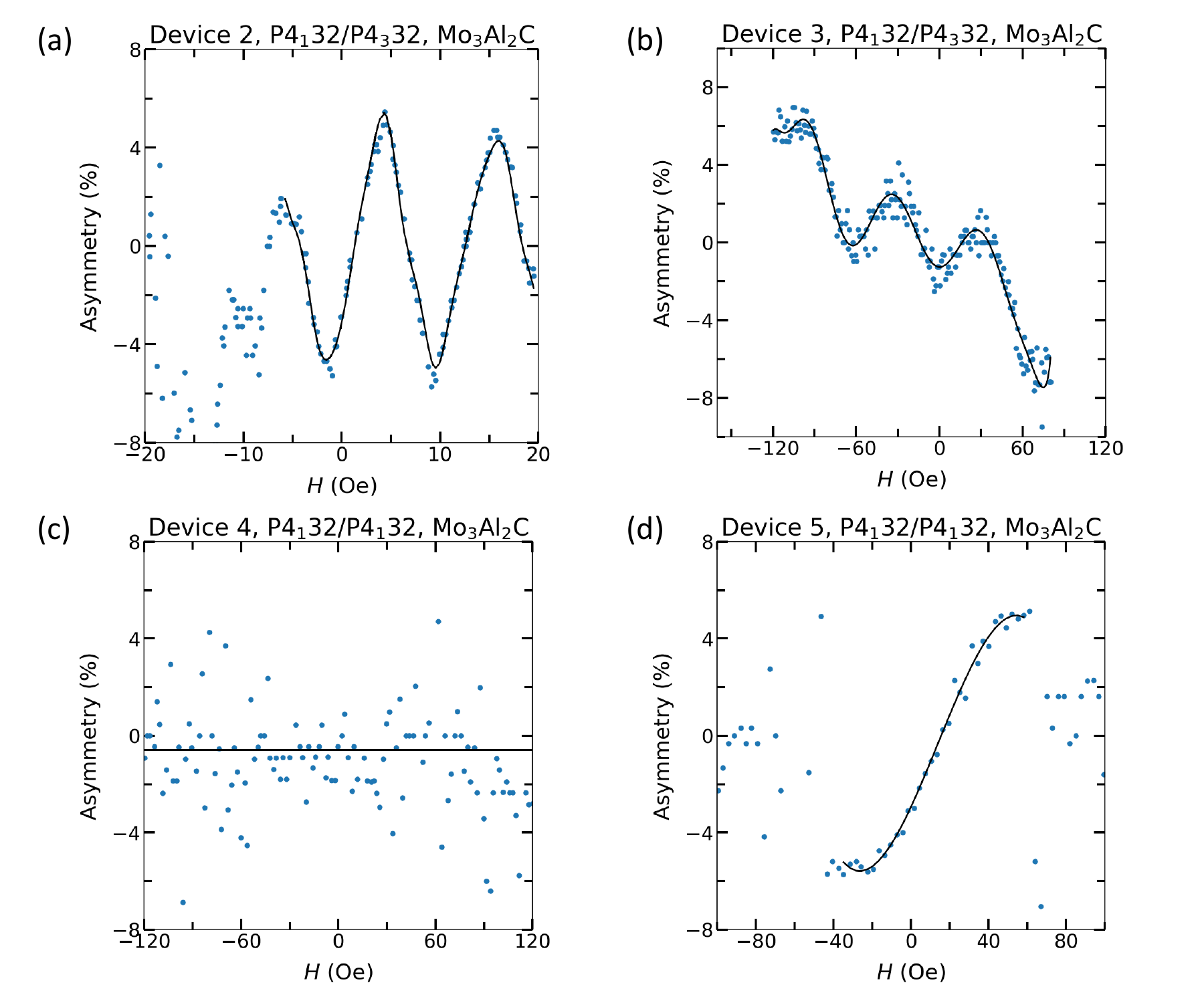}
\caption{Asymmetry versus applied magnetic field for (a) Device 2 (P4$_1$32/P4$_3$32), (b) Device 3 (P4$_1$32/P4$_3$32), (c) Device 4 (P4$_1$32/P4$_1$32), and (d) Device 5 (P4$_1$32/P4$_1$32). The horizontal dashed line represents where the asymmetry is 0 \%, and the vertical dashed line is an estimate of where true zero field is.}
\label{fig:enter-label}
\end{figure}

\begin{figure}
\centering
\includegraphics[width=1.2 \linewidth]{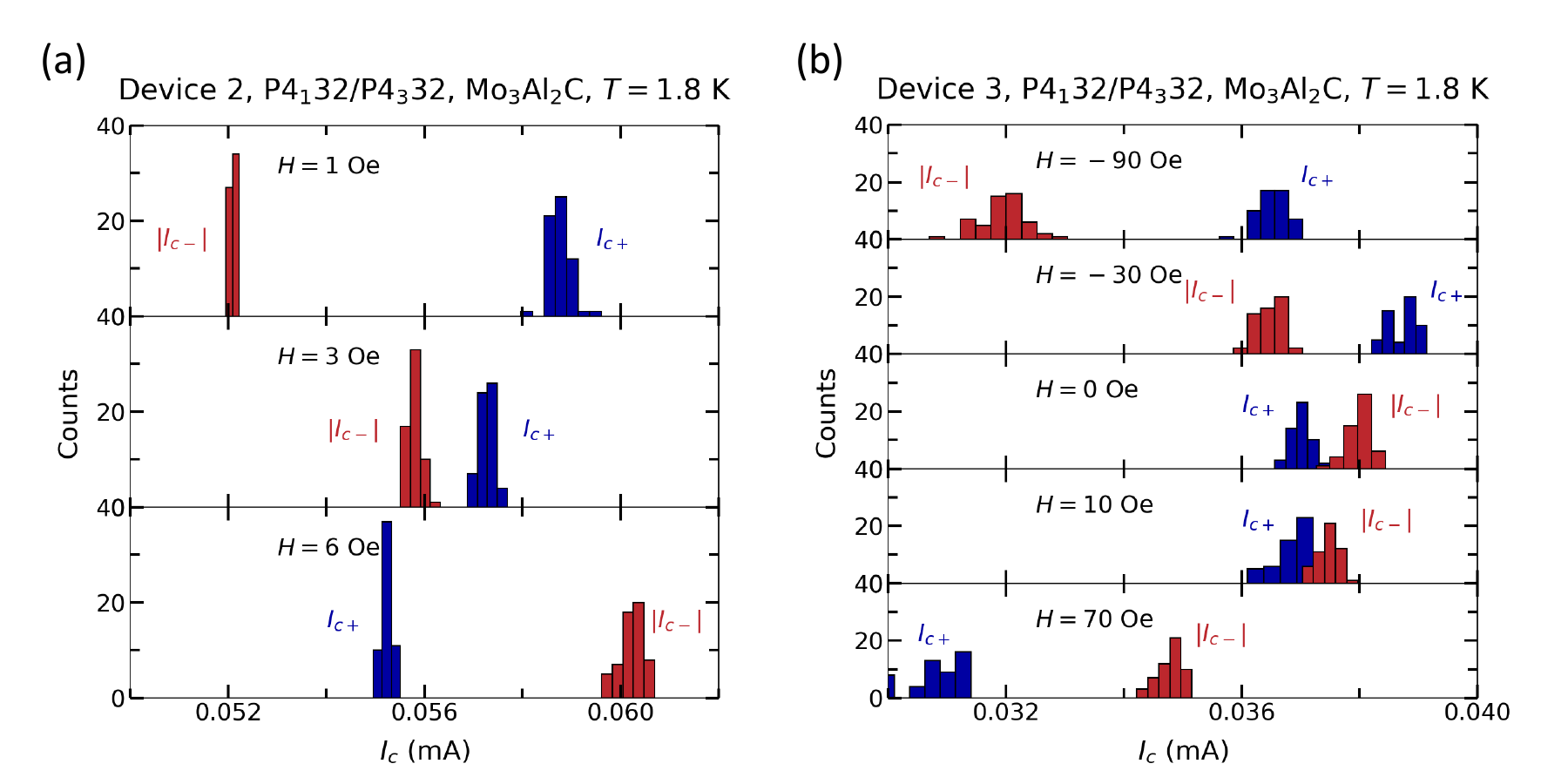}
\caption{Histogram plots of $I_{c+}$ and $|I_{c-}|$ at $T = 1.8$~K for applied magnetic fields of (a) $H = 1$~Oe, $H = 3$~Oe, and $H = 6$~Oe for Device 2 (P4$_1$32/P4$_3$32) and (b)$H=-90$~Oe, $H=-30$~Oe, $H=0$~Oe, $H=10$~Oe, and $H=70$~Oe for Device 3 (P4$_1$32/P4$_3$32).}
\label{fig:enter-label}
\end{figure}

\begin{figure}
\centering
\includegraphics[width= \linewidth]{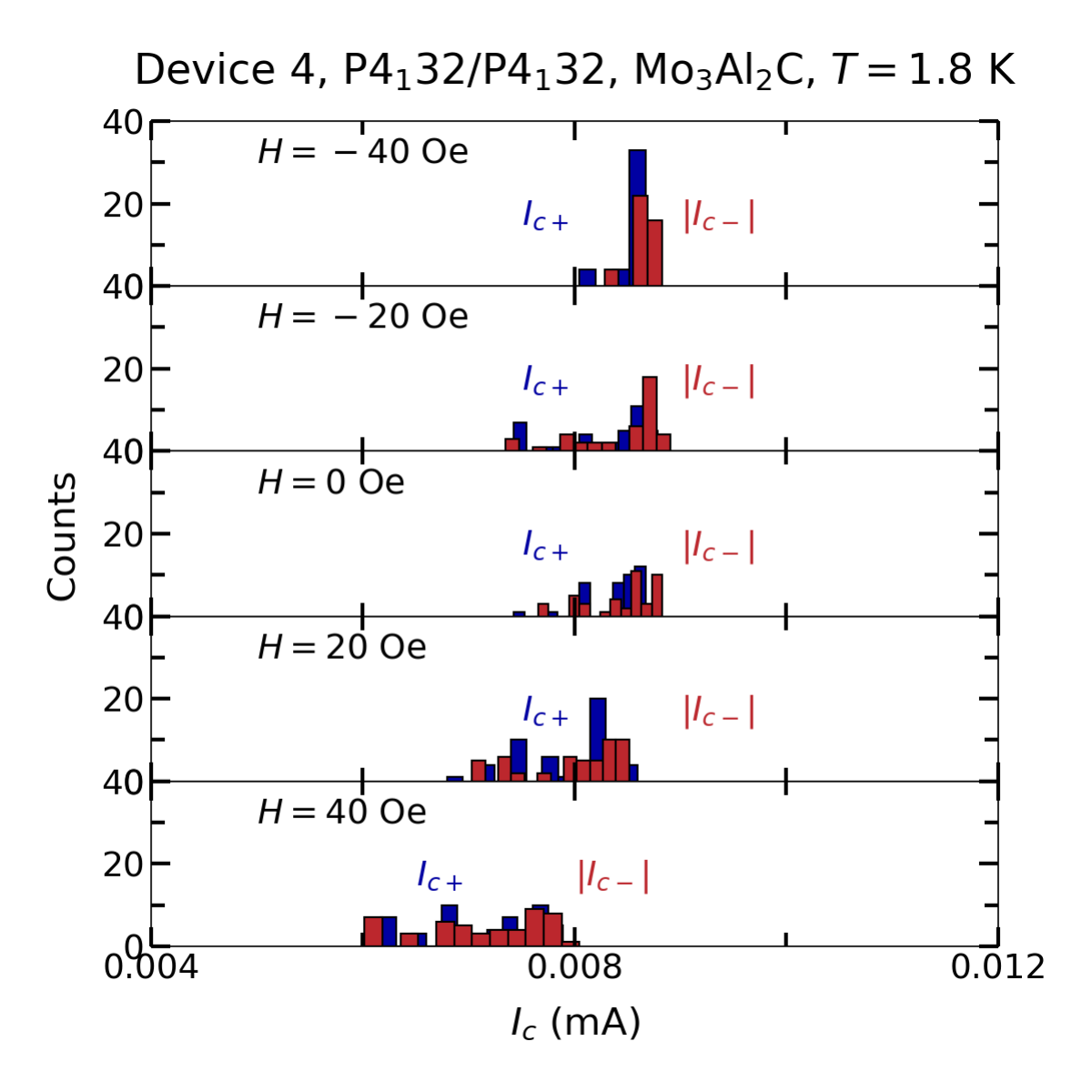}
\caption{Histogram plots of $I_{c+}$ and $|I_{c-}|$ at $T = 1.8$~K for applied magnetic fields of $H = -40$~Oe, $H = -20$~Oe, $H = 0$~Oe, $H = 20$~Oe, and $H = 40$~Oe  for Device 4 (P4$_1$32/P4$_1$32).}
\label{fig:enter-label}
\end{figure}

\clearpage
\renewcommand{\figurename}{SI Figure}
\renewcommand{\tablename}{SI Table}
\setcounter{figure}{0}
\counterwithout{figure}{section}
\counterwithout{table}{section}
\begin{titlepage}
    \centering
    \maketitle
    {\LARGE\bfseries Supporting Information: The superconducting diode effect in Josephson junctions fabricated from structurally chiral Mo$_3$Al$_2$C}
    
\end{titlepage}

\section{X-ray Diffraction}
\subsection{Powder X-ray Diffraction}
\begin{figure}[H]
\centering
\includegraphics[width= 1 \linewidth]{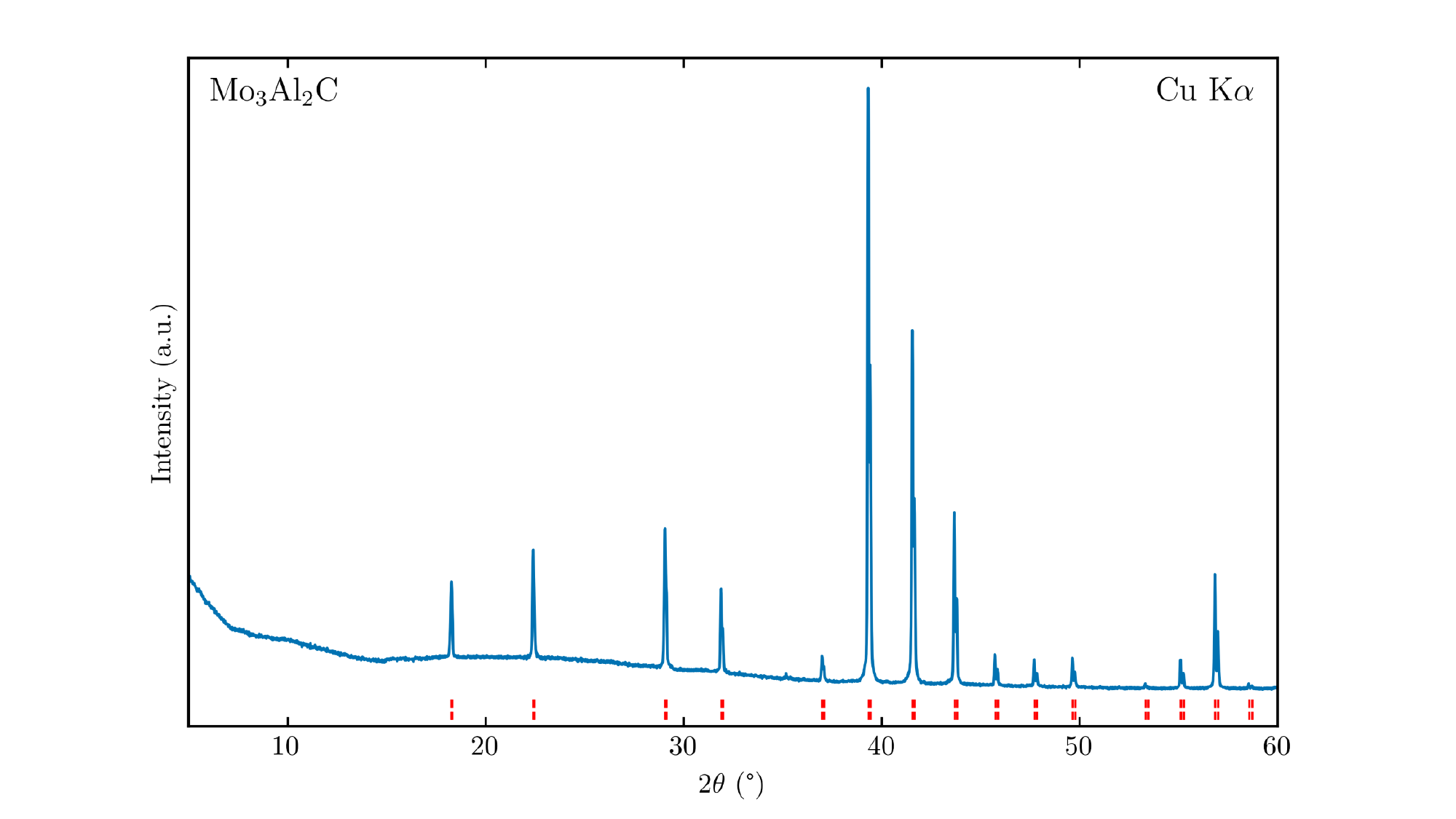}
\caption{Powder X-ray diffraction pattern of of ground metallic pieces of Mo$_3$Al$_2$C.}
\end{figure}

\subsection{Single Crystal X-ray Diffraction}
\subsubsection{Single Crystal X-ray Diffraction Refinement Data}
\begin{table}[h]
\caption{Mo$_3$Al$_2$C single crystal X-ray diffraction refinement data.}\label{tab1}%
\begin{tabular}{|p{5 cm}|p{4 cm}|p{4 cm}|}
\hline
Mo$_3$Al$_2$C & Crystal 1 & Crystal 2 \\
\hline
Crystal System   & Cubic   & Cubic  \\
Space Group    & P4$_3$32   & P4$_1$32  \\
$a$ (\AA)   & 6.85503(9)   & 6.85972(9)  \\
Volume (\AA$^{3}$)    & 322.127(7)   & 322.789(13) \\
$Z$   & 2   & 2  \\
Radiation   & Mo $K \alpha$ ($\lambda = 0.71073$ \AA)   & Mo $K \alpha$ ($\lambda = 0.71073$ \AA)  \\
Temperature (K)    & 213(2)   & 213(2)  \\
No. of measured, independent, and observed [$I > 2 \sigma I$] reflections  & 9857, 213, 212   & 9880, 213, 212  \\
$R_{int}$    & 0.0358   & 0.0355  \\
Goodness-of-fit   & 1.143   & 1.164  \\
$R_1 [F^{2} > 2 \sigma (F^{2})]; R_1 $[all data]    & 0.0069; 0.0070   & 0.0080; 0.0082   \\
$wR(F^{2})$   & 0.0167   & 0.196  \\
Largest diff peak and hole (e \AA$^{-3}$)   & 0.342 and -0.288   & 0.434 and -0.384  \\
Flack parameter, x    & -0.03(4)   & -0.08(4)  \\\botrule
\end{tabular}

\end{table}

\begin{table}
\caption{Refined atomic positions and anisotropic displacement parameters for Mo$_3$Al$_2$C, crystal 1 and crystal 2. }\label{tab1}%
\begin{tabular}{ |p{2cm}|p{1.5cm}|p{1.5cm}|p{1.5cm}|p{1.5cm}|p{1.5cm}|p{1.5cm}|}
\hline
 & \multicolumn{3}{|c|}{Crystal 1} & \multicolumn{3}{|c|}{Crystal 2} \\
\hline
Atom & Mo & Al & C & Mo & Al & C \\
\hline
Occupancy & 1.00 & 1.00 & 1.00 & 1.00 & 1.00 & 1.00 \\
Wyckoff Site & 12d & 8c   & 4a & 12d & 8c & 4a \\
$x$ & 0.375000  & 0.43260(7) & 0.125000 & 0.625000 & 0.56744(7) & 0.875000 \\
$y$ & 0.29731(2)   & 0.43260(7) & 0.125000 & 0.70267(2) & 0.56744(7) & 0.875000 \\
$z$ & 0.04731(2) & 0.43260(7) & 0.125000 & 0.95267(2) & 0.56744(7) & 0.875000 \\
$U_{11}$ & 0.00835(10) & 0.00213(15) & 0.0062(6) & 0.00916(11) & 0.00304(17) & 0.0069(7)   \\
$U_{22}$ & 0.00330(7) & 0.00213(15) & 0.0062(6) & 0.00418(8) & 0.00304(17) & 0.0069(7) \\
$U_{33}$ & 0.00330(7) & 0.00213(15) & 0.0062(6) & 0.00418(8) & 0.00304(17) & 0.0069(7) \\
$U_{23}$ & -0.00038(6) & 0.00021(15) & -0.0004(8) & -0.00029(7) & 0.00030(17) & -0.0010(8) \\
$U_{13}$ & 0.00175(5) & 0.00021(15) & -0.0004(8) & 0.00178(5) & 0.00030(17) & -0.0010(8) \\
$U_{12}$ & -0.00175(5) & 0.00021(15) & -0.0004(8) & -0.00178(5) & 0.00030(17) & -0.0010(8) \\
\hline
\end{tabular}
\end{table}

\newpage
\subsubsection{Handedness Determination for Devices}

To sort left and right-handed crystals, abbreviated data sets were taken to collect the redundant data, $(hkl)$ and $(-h,-k,-l)$ reflections (Friedel pairs), to determine the handedness of each crystal. A summary of the crystals used for the devices is shown in SI Table 3. We ran the refinement assuming a P4$_3$32 crystal structure, so $x = 0$ and $x = 1$ correspond to the P4$_3$32 and P4$_1$32 crystal structures, respectively. 

\begin{table}[h]
\caption{A summary of the crystals and their Flack parameters used to fabricate the devices.}\label{tab1}%
\begin{tabular}{@{}llll@{}}
\toprule
Device Number & Flack Parameter for Crystal 1 & Flack Parameter for Crystal 2 & Device Type \\
\midrule
1    &  0.74(13)  & N/A  & Single Crystal \\
2   &  0.93(13)  & -0.15(14) &  P4$_1$32/P4$_3$32 \\
3   &  0.96(13)  & 0.03(10) &  P4$_1$32/P4$_3$32 \\
4    &  1.07(10)  & 1.18(16) & P4$_1$32/P4$_1$32 \\
5    &  0.91(16)  & 1.03(12) & P4$_1$32/P4$_1$32 \\
6  & 0.85(15) & 0.12(10) & P4$_1$32/P4$_3$32  \\
7   &  0.88(11)  & 0.96(12) &  P4$_1$32/P4$_1$32 \\
8    &  1.00(12)  & 1.08(15) & P4$_1$32/P4$_1$32 \\\botrule
\end{tabular}

\end{table}

\section{Physical Properties Measurements}
\subsection{Magnetization Measurements}

\begin{figure}[H]
\centering
\includegraphics[width= 0.8 \linewidth]{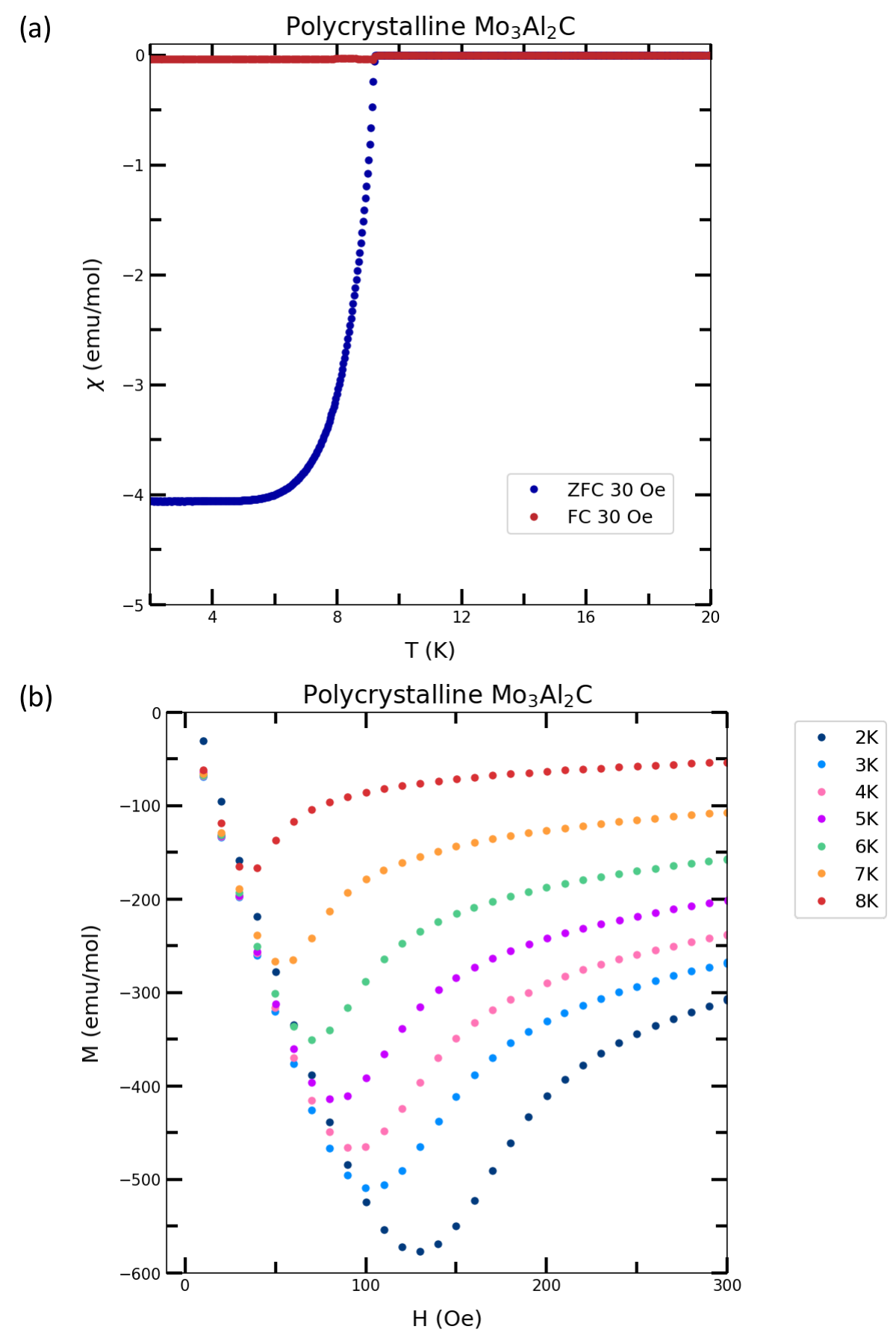}
\caption{(a) Magnetic susceptibility versus temperature and (b) magnetization versus magnetic field for a polycrystalline sample of Mo$_3$Al$_2$C. A superconducting transition is observed at $T_{c} = 8.5$~K, and the critical field is estimated to be $H~=~120$~Oe at $T = 2$~K. }
\end{figure}

\subsection{Resistance Measurements}
\begin{figure}[!htb]
\centering
\includegraphics[width=\linewidth]{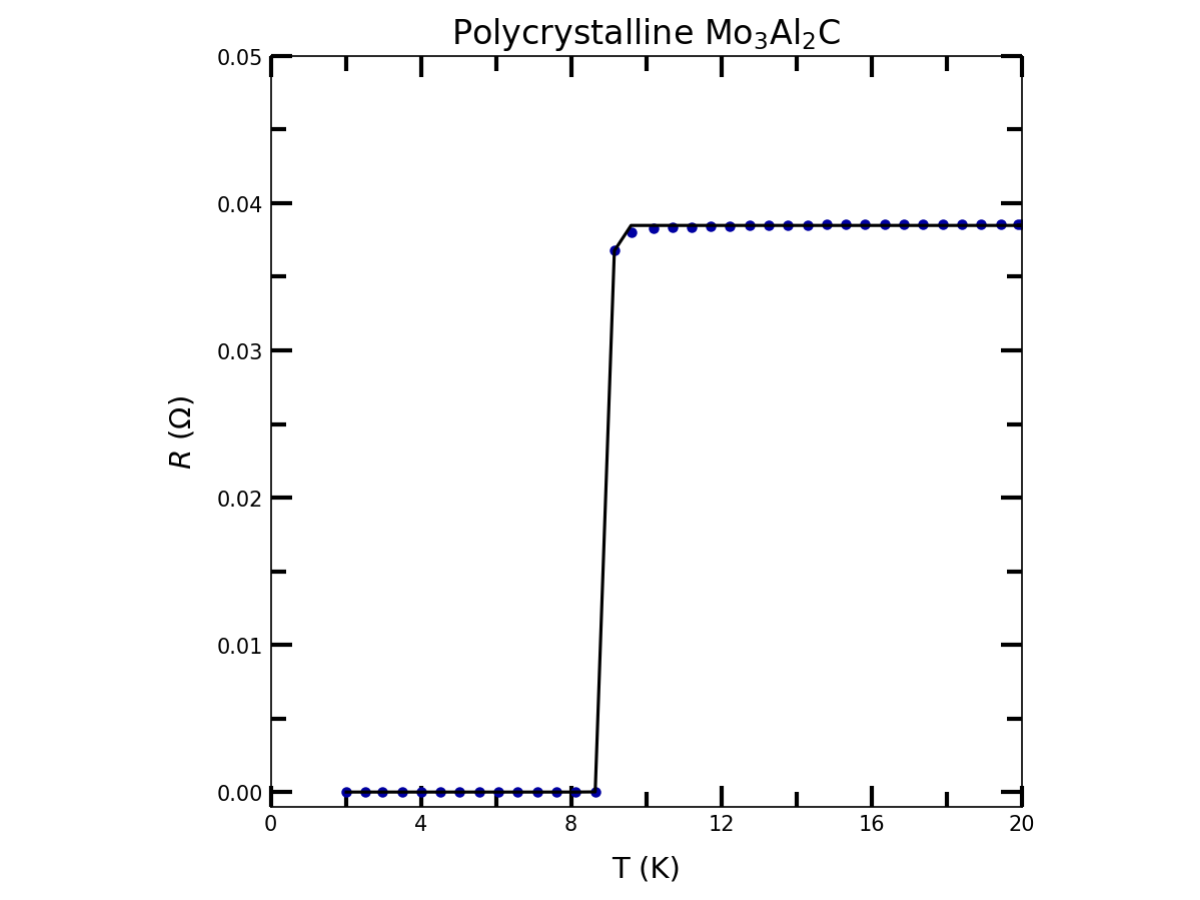}
\caption{Four-probe resistance versus temperature for a polycrystalline sample of Mo$_3$Al$_2$C. A sharp superconducting transition is observed at $T_{c} = 8.99(10)$~K.}
\end{figure}

\section{Device Characterization}
\subsection{Temperature Dependence of IV Curves}
\begin{figure}[H]
\centering
\includegraphics[width=\linewidth]{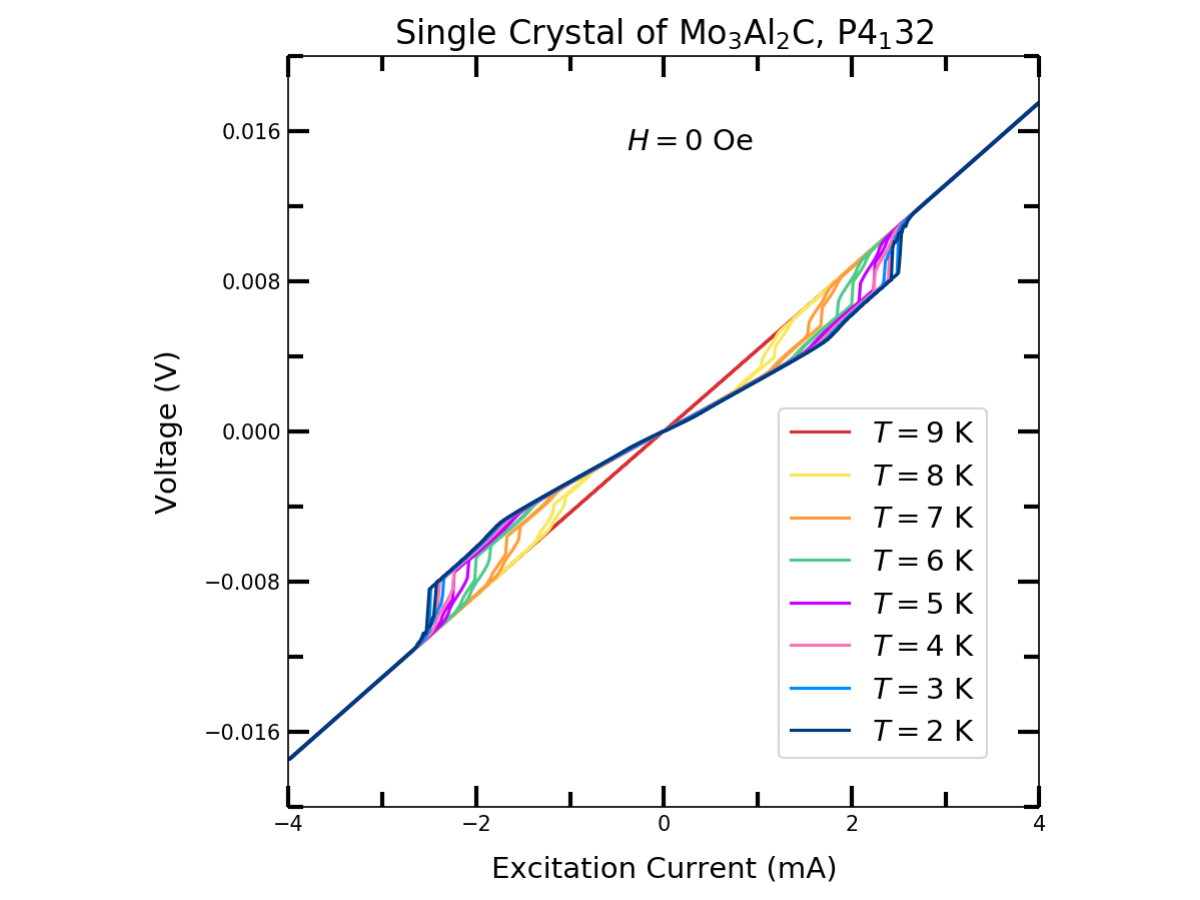}
\caption{IV curves at temperatures from $T = 9$~K to $T = 2$~K for the single crystal device. Below $T_c$, a critical current is observed.}
\end{figure}

\begin{figure}[H]
\centering
\includegraphics[width= 0.9 \linewidth]{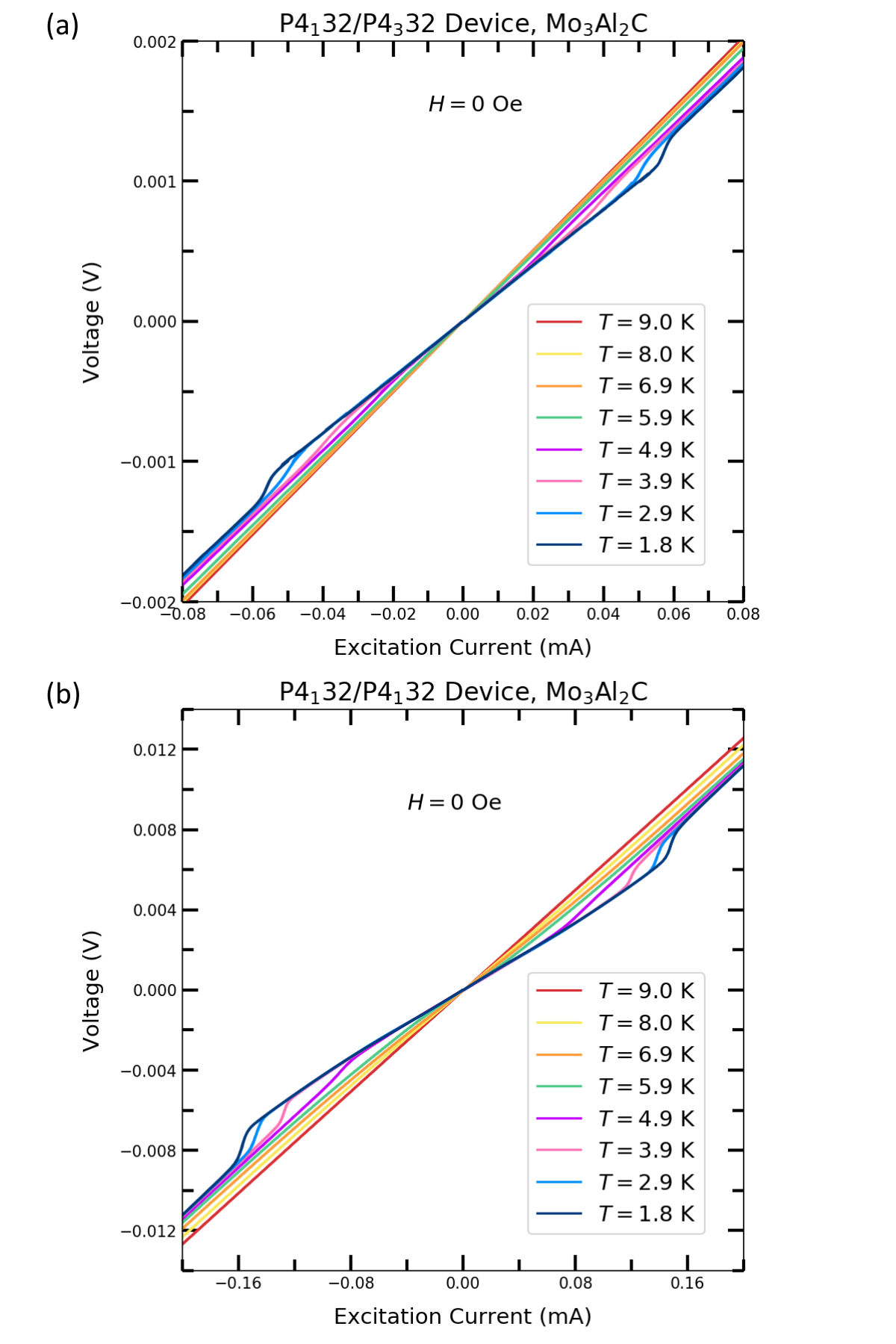}
\caption{IV curves at temperatures from $T = 9$~K to $T = 1.8$~K for the (a) P4$_1$32/P4$_3$32 (device 2) and (b)  P4$_1$32/P4$_1$32 (device 5) devices. The critical current is not reliably observable until $T << T_{c}$, i.e. $T \approx 3$~K for the P4$_1$32/P4$_3$32 device and $T \approx 4$~K for the P4$_1$32/P4$_1$32 device.}
\end{figure}

\subsection{Forward and Reverse Sweep IV Curves}
\begin{figure}[!htb]
\centering
\includegraphics[width=\linewidth]{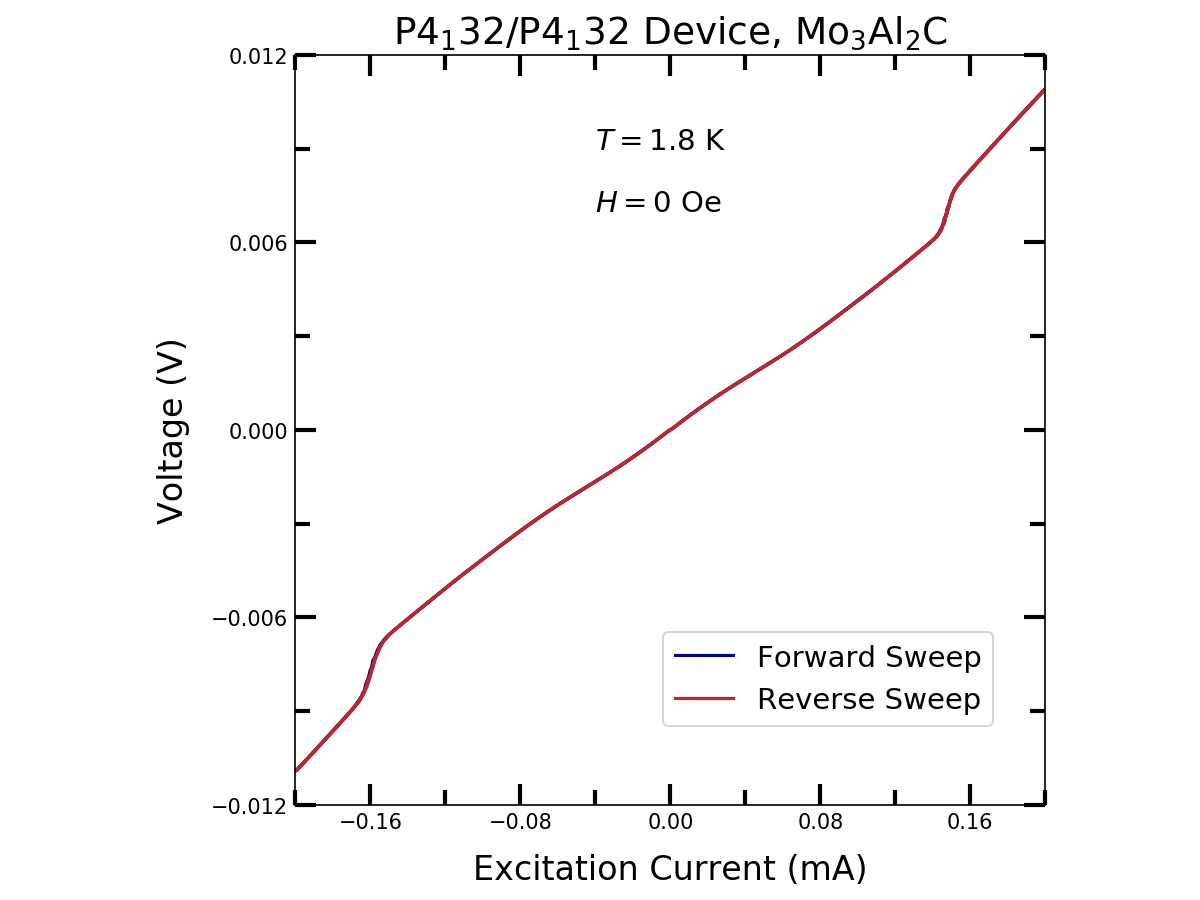}
\caption{IV Curve of the P4$_1$32/P4$_1$32 device (device 5) with the current swept in the forward ($0$ to $I_{\text{max}}$, $I_{\text{max}}$ to $0$, $0$ to $-I_{\text{max}}$, and $-I_{\text{max}}$ to $0$) and reserve ($0$ to $-I_{\text{max}}$, $-I_{\text{max}}$ to $0$, $0$ to $I_{\text{max}}$, and $I_{\text{max}}$ to $0$) direction.}
\end{figure}

\subsection{Magnetic Field Dependence of the Critical Current}
\begin{figure}[H]
\centering
\includegraphics[width=0.8 \linewidth]{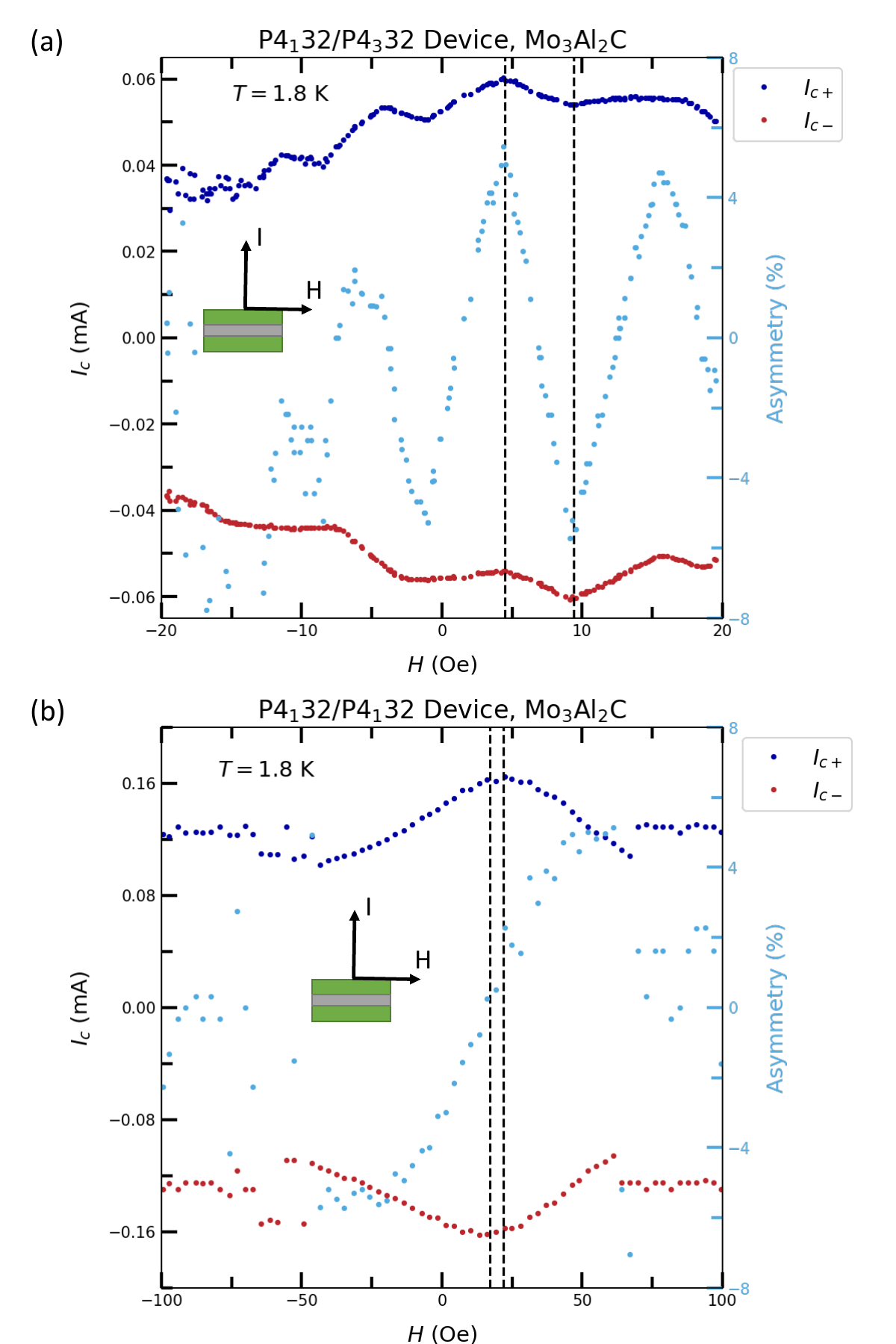}
\caption{Critical current and asymmetry versus magnetic field for (a) the P4$_1$32/P4$_3$32 device (device 2) and (b) the P4$_1$32/P4$_1$32 device (device 5). The magnetic field is applied in the plane of the insulating barrier.}
\end{figure}

\begin{figure}[H]
\centering
\includegraphics[width=\linewidth]{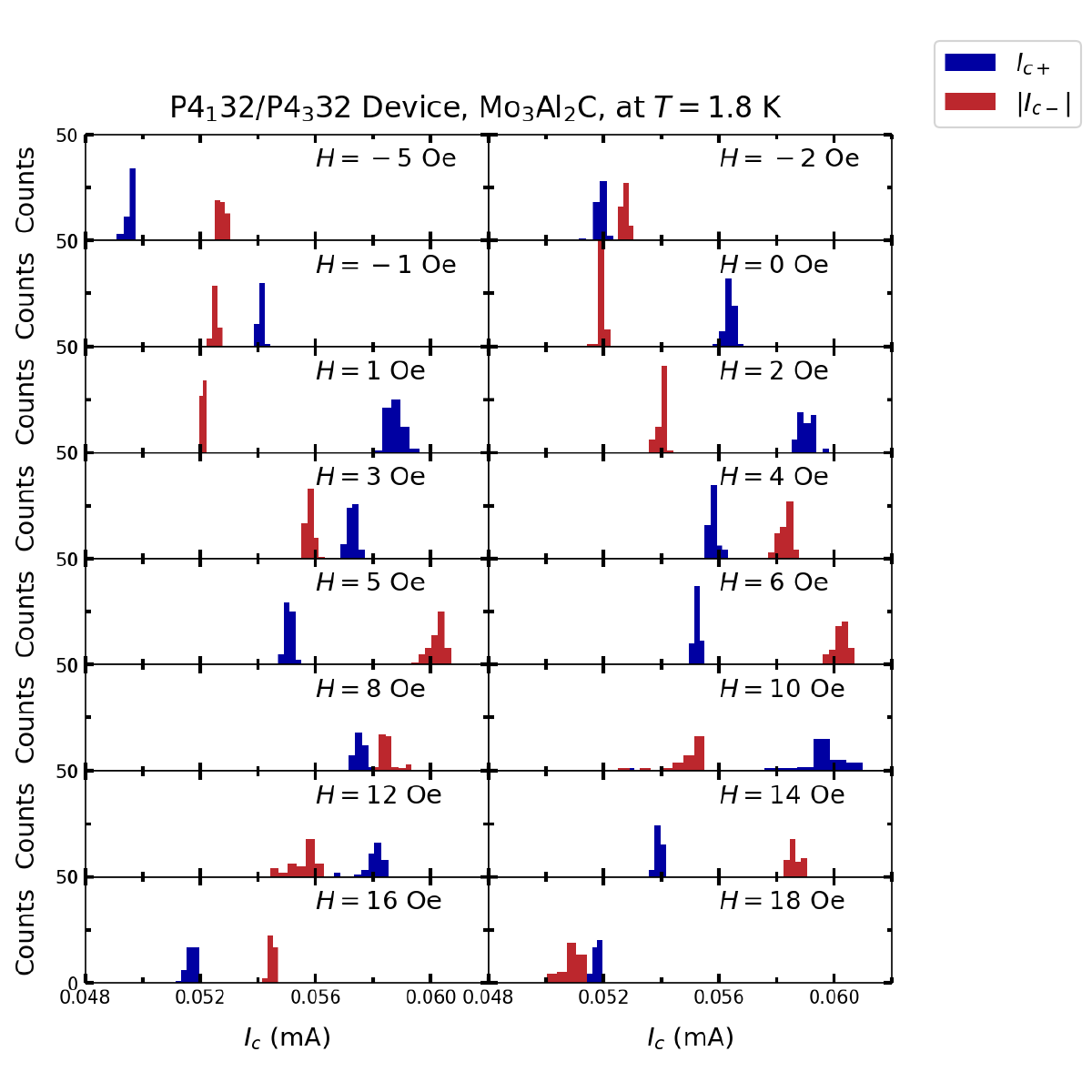}
\caption{Histogram plots for the P4$_1$32/P4$_3$32 device (device 2) of $I_{c}$ at magnetic fields of $H = -5$~Oe to $H = 18$~Oe showing how the diode direction switches as a function of magnetic field. }
\end{figure}

\begin{figure}[H]
\centering
\includegraphics[width=\linewidth]{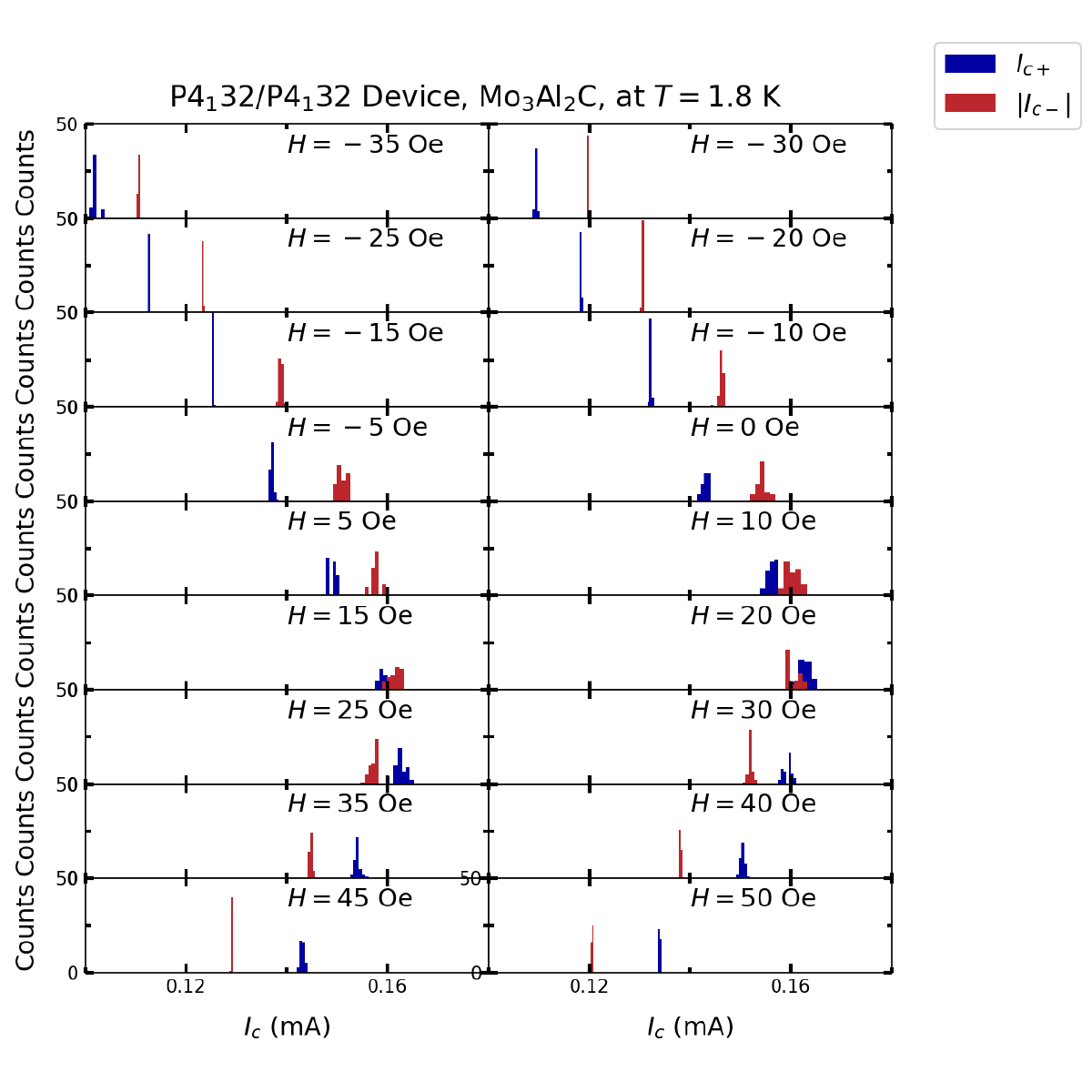}
\caption{Histogram plots for the P4$_1$32/P4$_1$32 device (device 5) of $I_{c}$ at magnetic fields of $H = -35$~Oe to $H = 50$~Oe showing how the diode direction switches as a function of magnetic field.}
\end{figure}

The statistical significance is calculated to characterize the difference between $I_{c+}$ and $|I_{c-}|$ as
\begin{equation}
    t = \frac{\big|I_{c+} - |I_{c-}| \big|}{\sigma} = \frac{\big|I_{c+} - |I_{c-}| \big|}{\sqrt{\sigma(I_{c+})^2 + \sigma(I_{c-})^2}},
\end{equation}
where $\sigma(I_{c+})$ and $\sigma(I_{c-})$ are the standard deviations of the $I_{c+}$ and $|I_{c-}|$ distributions, respectively.

\subsection{Critical current versus magnetic field for forward and reverse sweeps}
\begin{figure}[H]
\centering
\includegraphics[width=1 \linewidth]{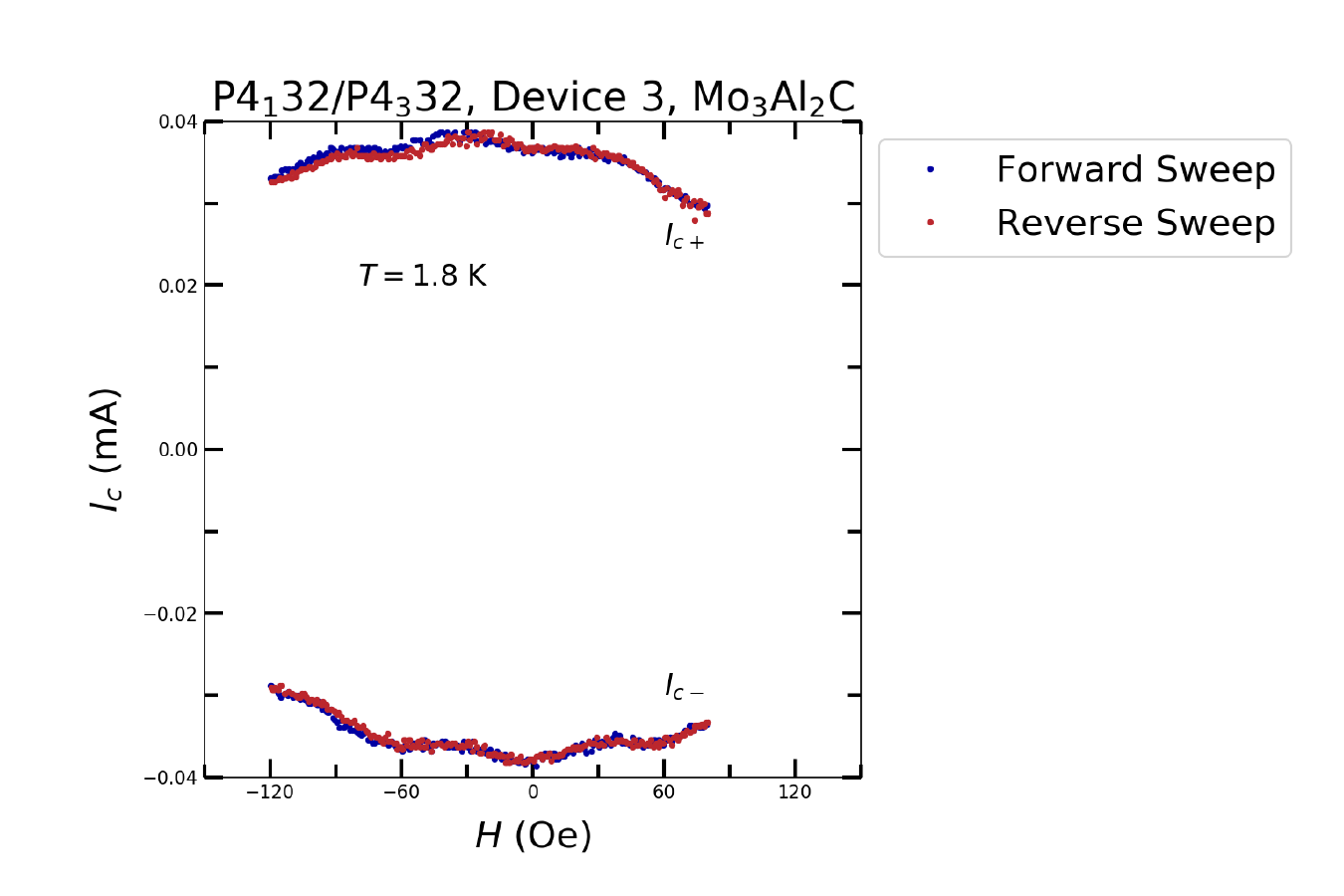}
\caption{Critical current versus magnetic field for the magnetic field swept in the forward (-120 Oe to 80 Oe) and reverse (80 Oe to -120 Oe) direction for device 3.}
\end{figure}

\begin{figure}[H]
\centering
\includegraphics[width=1 \linewidth]{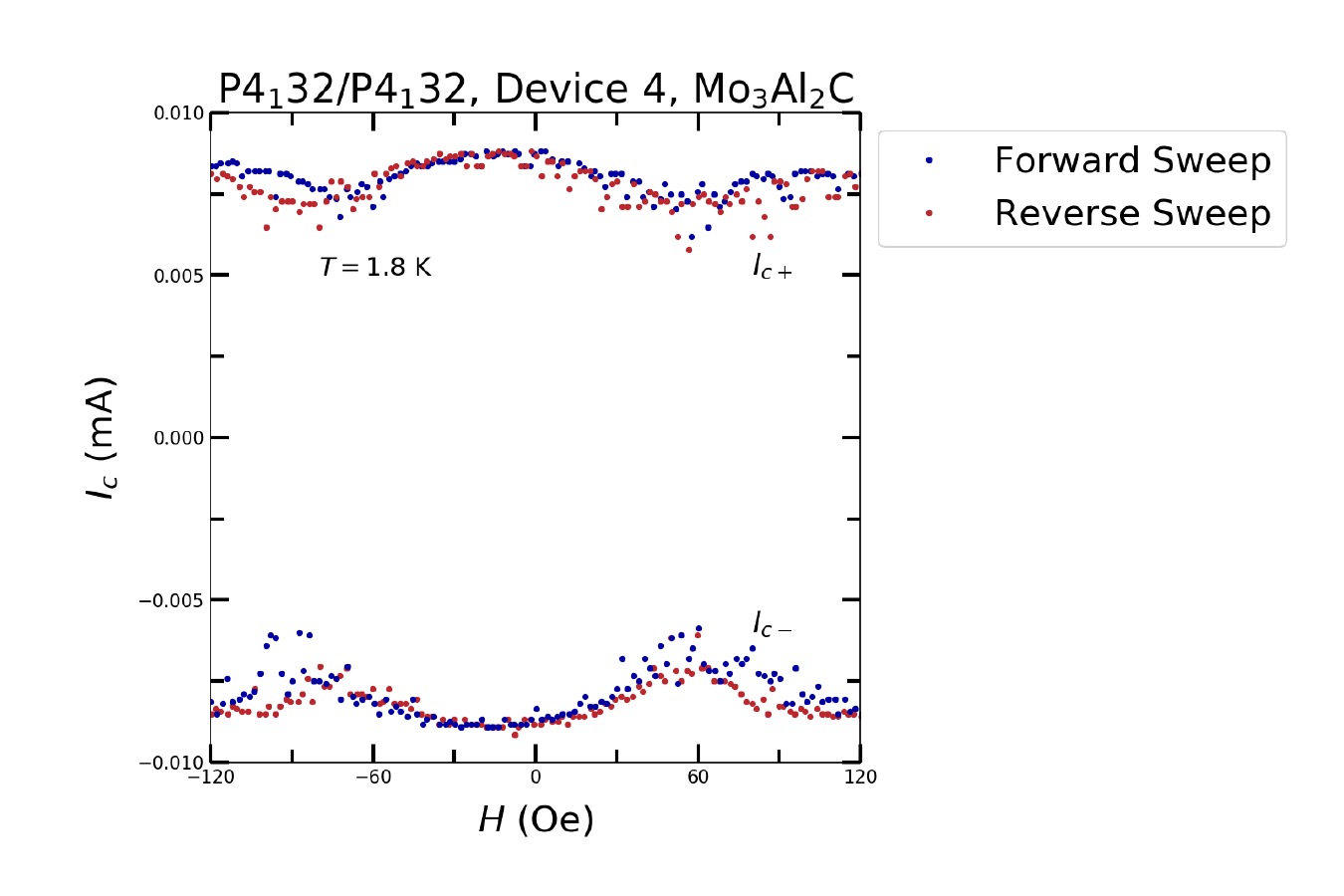}
\caption{Critical current versus magnetic field for the magnetic field swept in the forward (-120 Oe to 120 Oe) and reverse (120 Oe to -120 Oe) direction for device 4.}
\end{figure}

\begin{figure}[H]
\centering
\includegraphics[width=1 \linewidth]{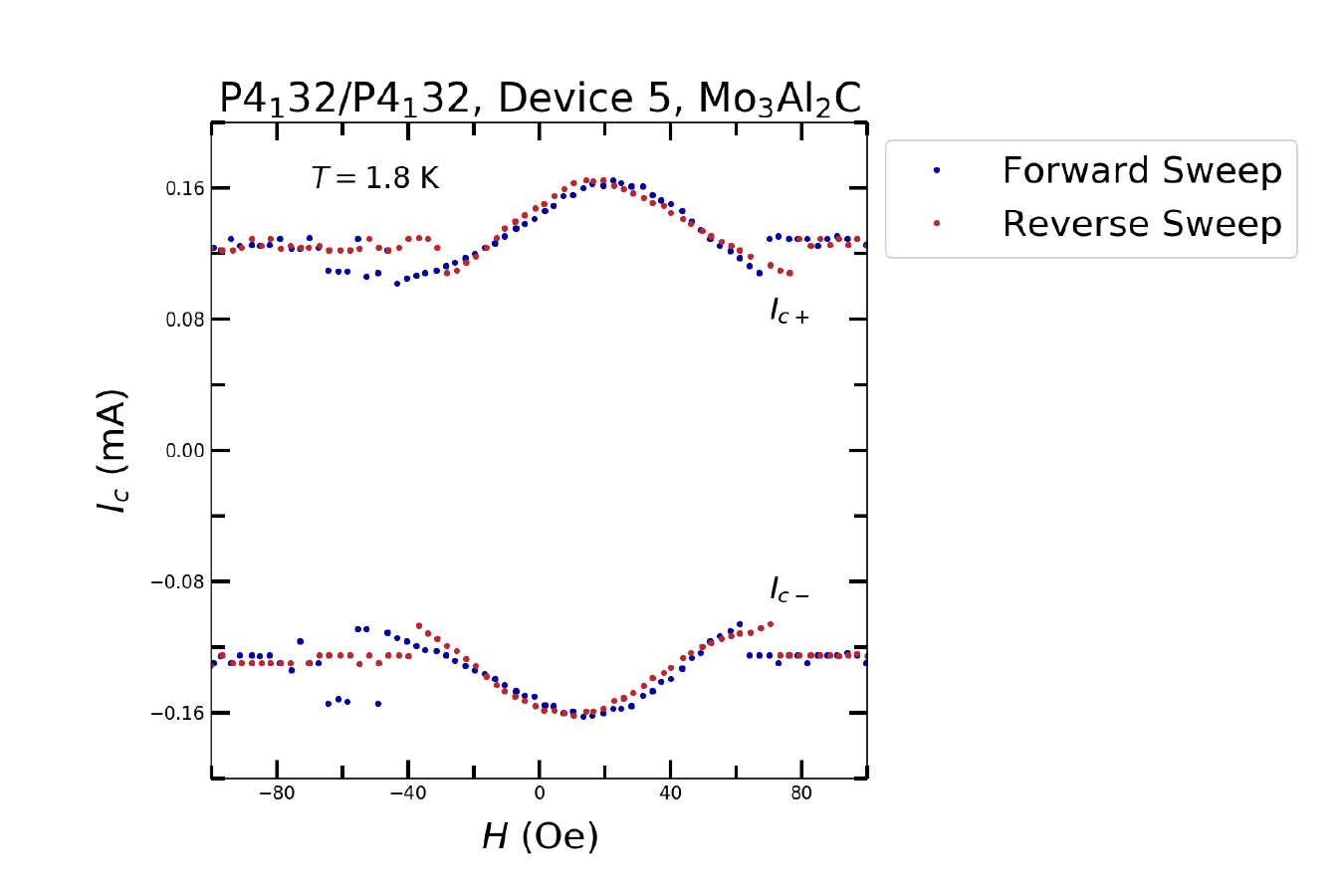}
\caption{Critical current versus magnetic field for the magnetic field swept in the forward (-100 Oe to 100 Oe) and reverse (100 Oe to -100 Oe) direction for device 5.}
\end{figure}

\subsection{Angular Dependence Measurements}
\begin{figure}[H]
\centering
\includegraphics[width= 0.8 \linewidth]{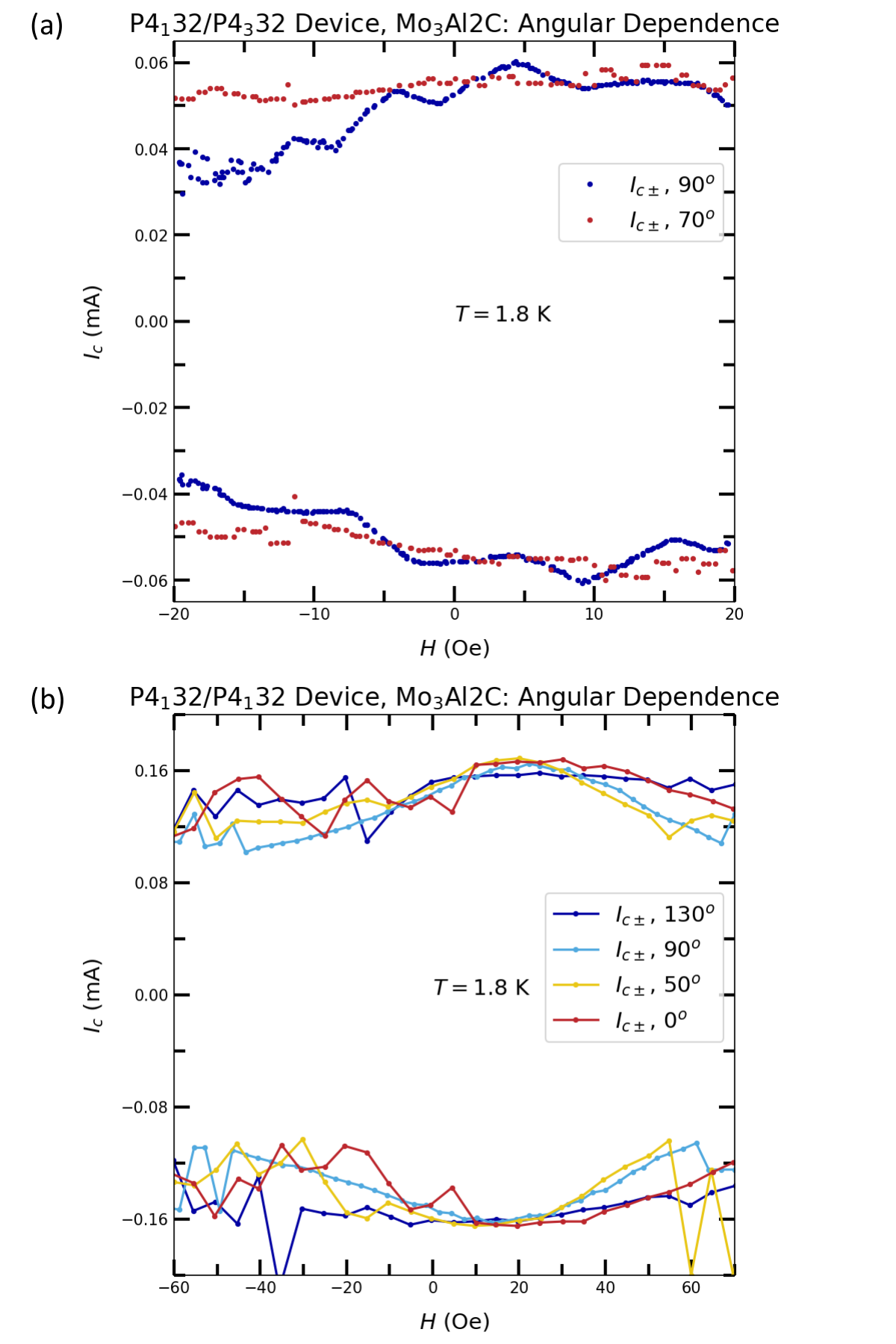}
\caption{(a) Critical current versus magnetic field for the P4$_1$32/P4$_3$32 device (device 2) with the rotator position at 90 degrees and 70 degrees and (b) critical current versus magnetic field for the P4$_1$32/P4$_1$32 device (device 5) with the rotator position at 130, 90, 50, and 0 degrees. The central peak broadens as the field is rotated out of the plane of the insulating barrier for device 2. Device 5 displays more complicated behavior, which is further evidence that it is in the wide junction limit.}
\end{figure}

\subsection{Additional Device Measurements}

\subsubsection{Device 6}
\begin{figure}[H]
\centering
\includegraphics[width= 0.8 \linewidth]{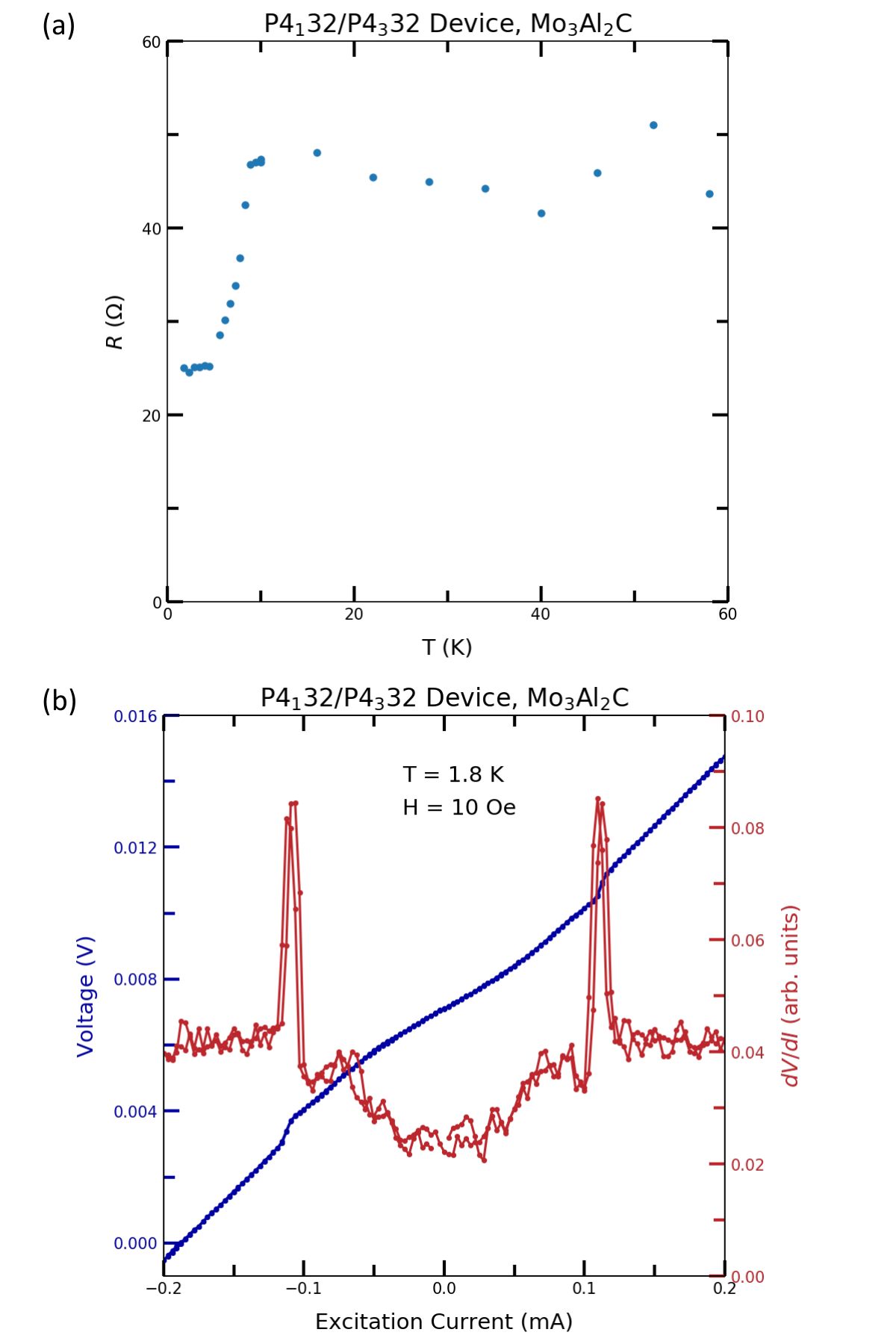}
\caption{(a) Resistance versus temperature and (b) voltage and $dV/dI$ versus excitation current for device 6.}
\end{figure}

\begin{figure}[H]
\centering
\includegraphics[width= 0.9 \linewidth]{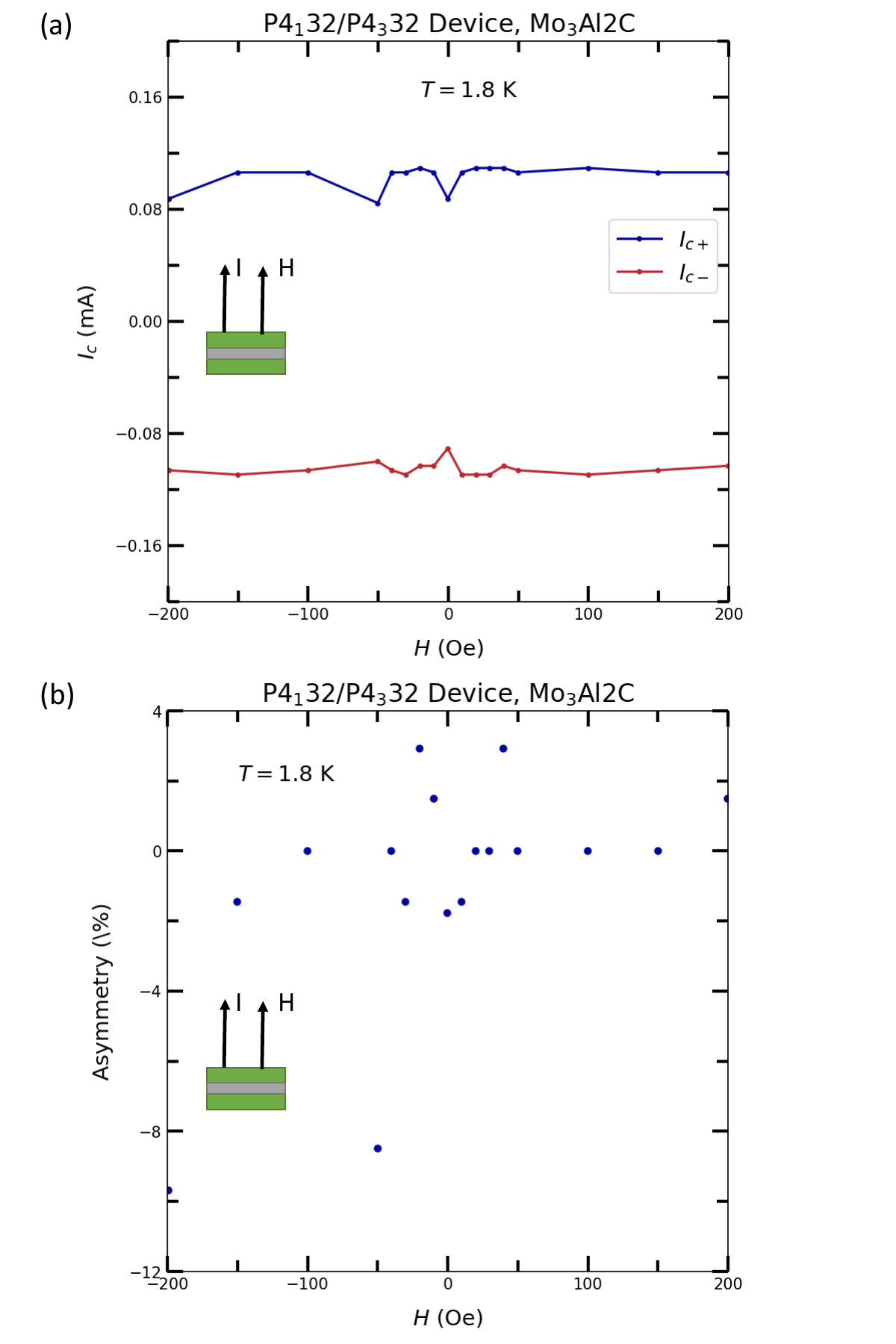}
\caption{(a) Critical current and (b) asymmetry versus magnetic field, where the field is applied perpendicular to the insulating barrier.}
\end{figure}

\subsubsection{Device 7}
\begin{figure}[H]
\centering
\includegraphics[width=\linewidth]{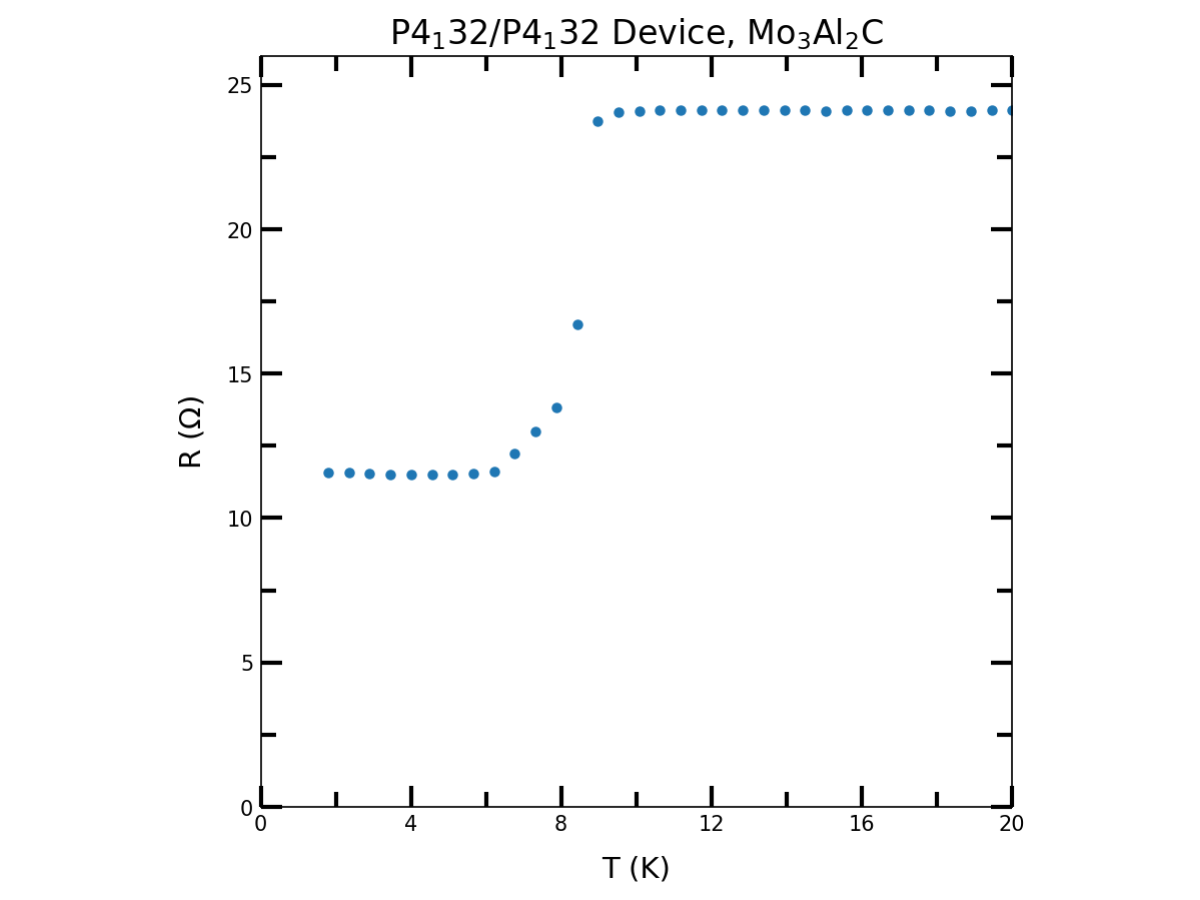}
\caption{Resistance versus temperature for P4$_1$32/P4$_1$32 device 7.}
\end{figure}

\begin{figure}[H]
\centering
\includegraphics[width= 0.85 \linewidth]{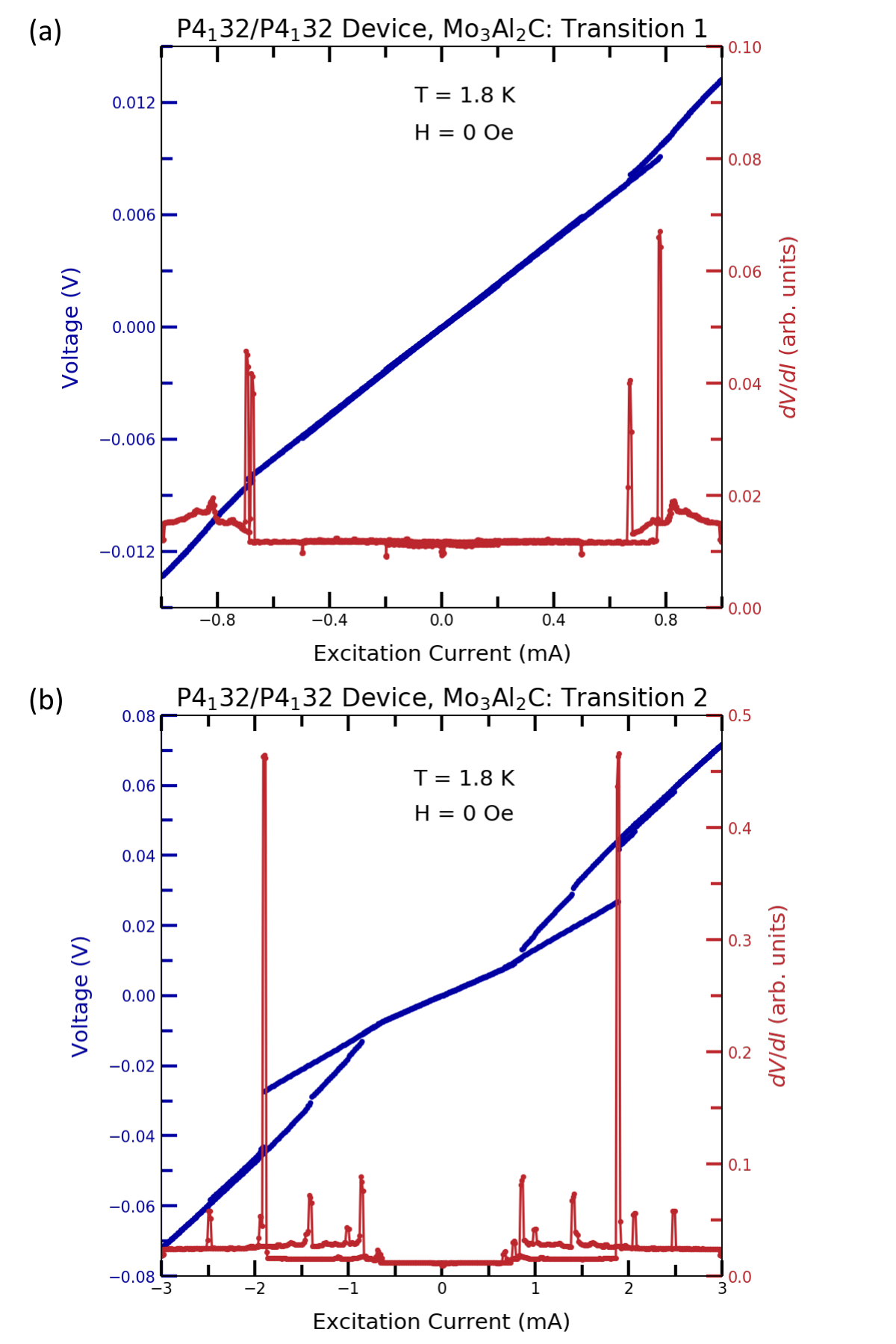}
\caption{IV curve and $dV/dI$ versus excitation current for the (a) first and (b) second transition observed in device 7. The second transition occurs at a much larger critical current than what was observed for other devices. We attribute this critical current directly to the superconductivity of Mo$_3$Al$_2$C, possibly due to misalignment of the crystals, allowing a superconducting pathway through one crystal. Therefore, the first transition observed, which is also significantly larger than the critical current observed in other man-made junctions, is possibly due to a grain boundary in the crystal, which forms a natural weak-link. }
\end{figure}

\begin{figure}[H]
\centering
\includegraphics[width= 0.9 \linewidth]{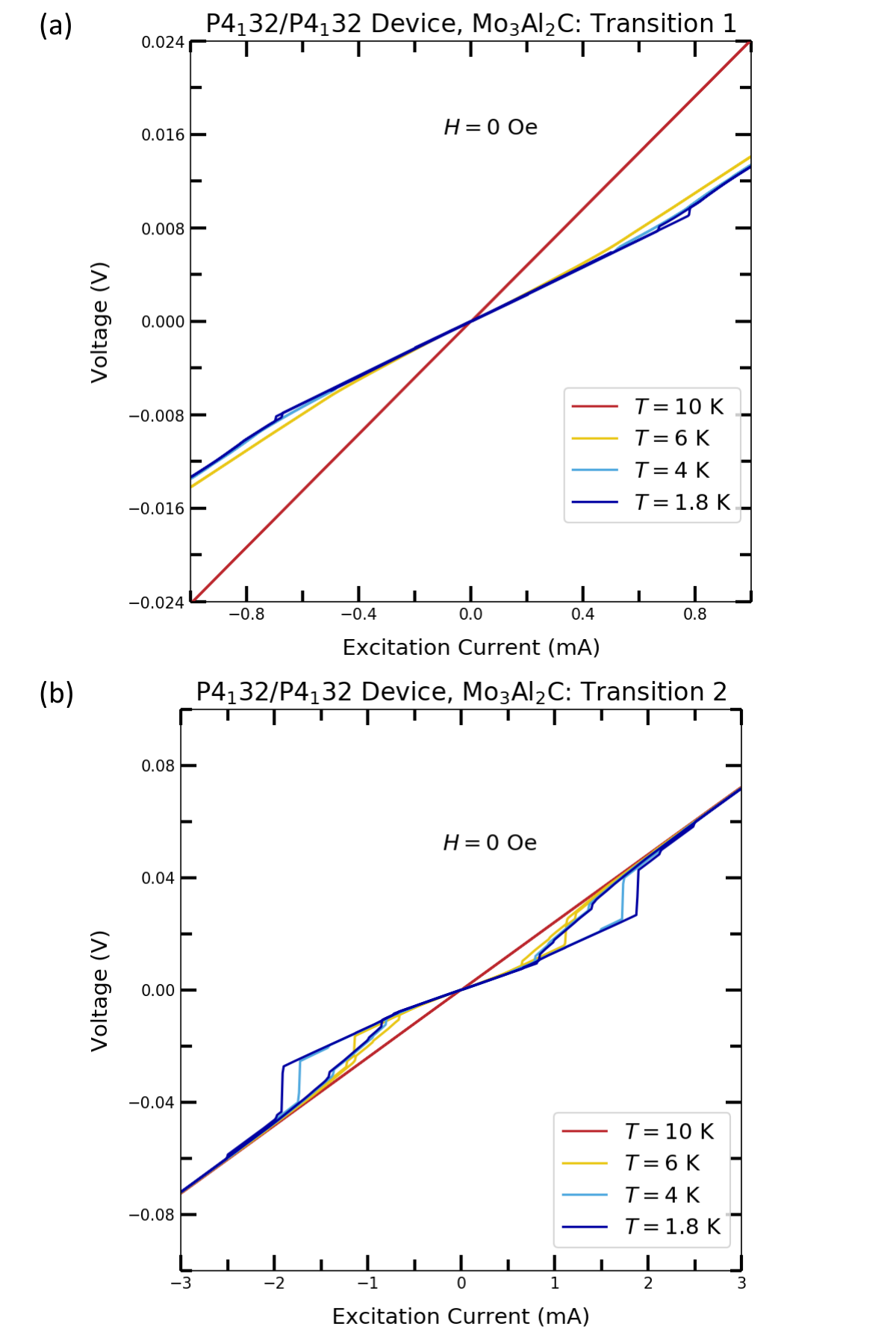}
\caption{IV curves at $T= 10, 6, 4,$ and $1.8$~K showing the temperature dependence of (a) transition 1 and (b) transition 2 in device 7.}
\end{figure}

\begin{figure}[H]
\centering
\includegraphics[width= 0.9 \linewidth]{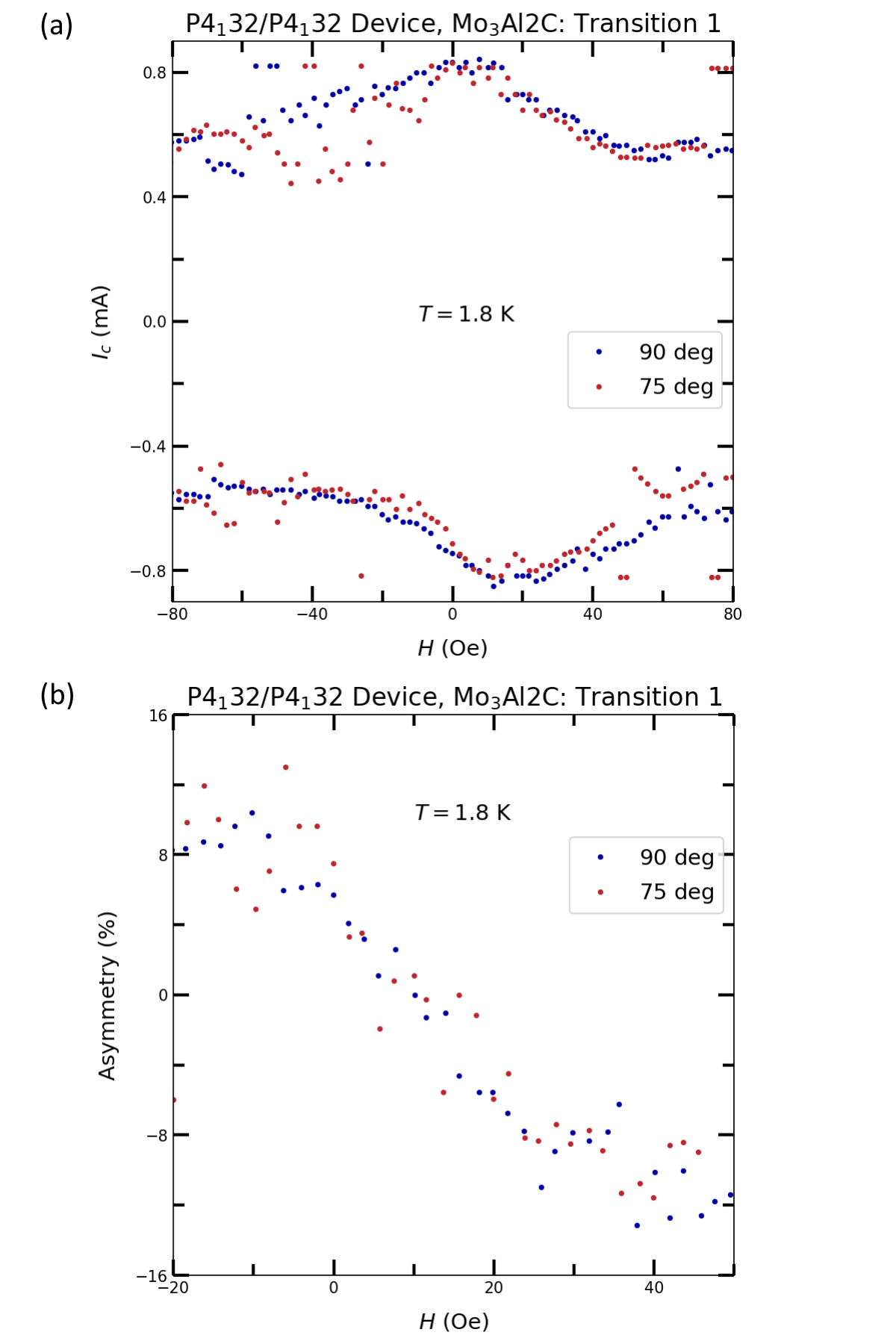}
\caption{(a) Critical current and (b) asymmetry versus magnetic field for the first transition in device 7 with the rotator position at 90 and 75 degrees. Since there are no oscillations in $I_c$ versus $H$, it is likely this device is in the wide junction limit.}
\end{figure}

\begin{figure}[H]
\centering
\includegraphics[width= 0.9 \linewidth]{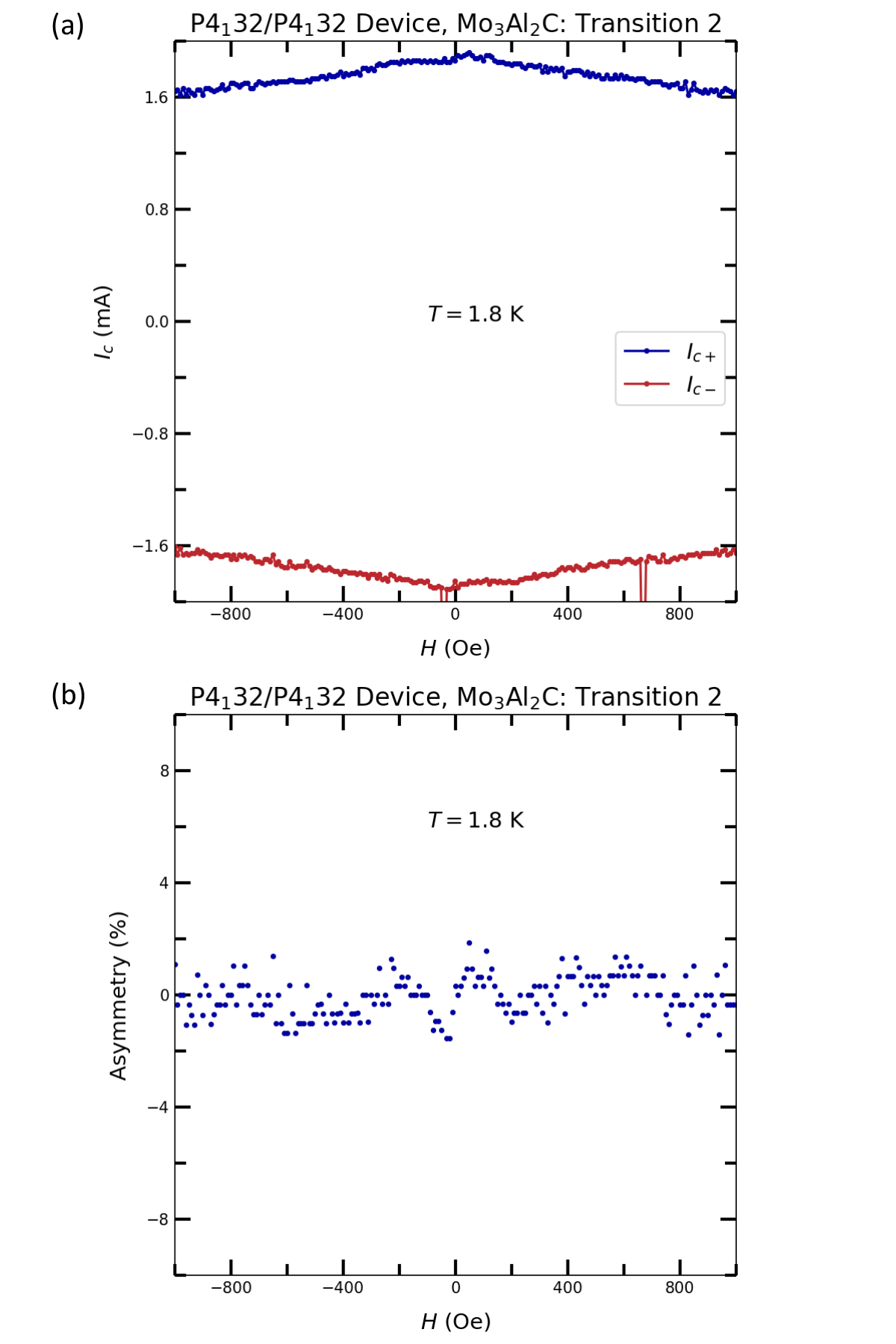}
\caption{(a) Critical current and (b) asymmetry versus magnetic field for the second transition in device 7 with the rotator position at 90 degrees.}
\end{figure}

\subsubsection{Device 8}
\begin{figure}[H]
\centering
\includegraphics[width=0.8 \linewidth]{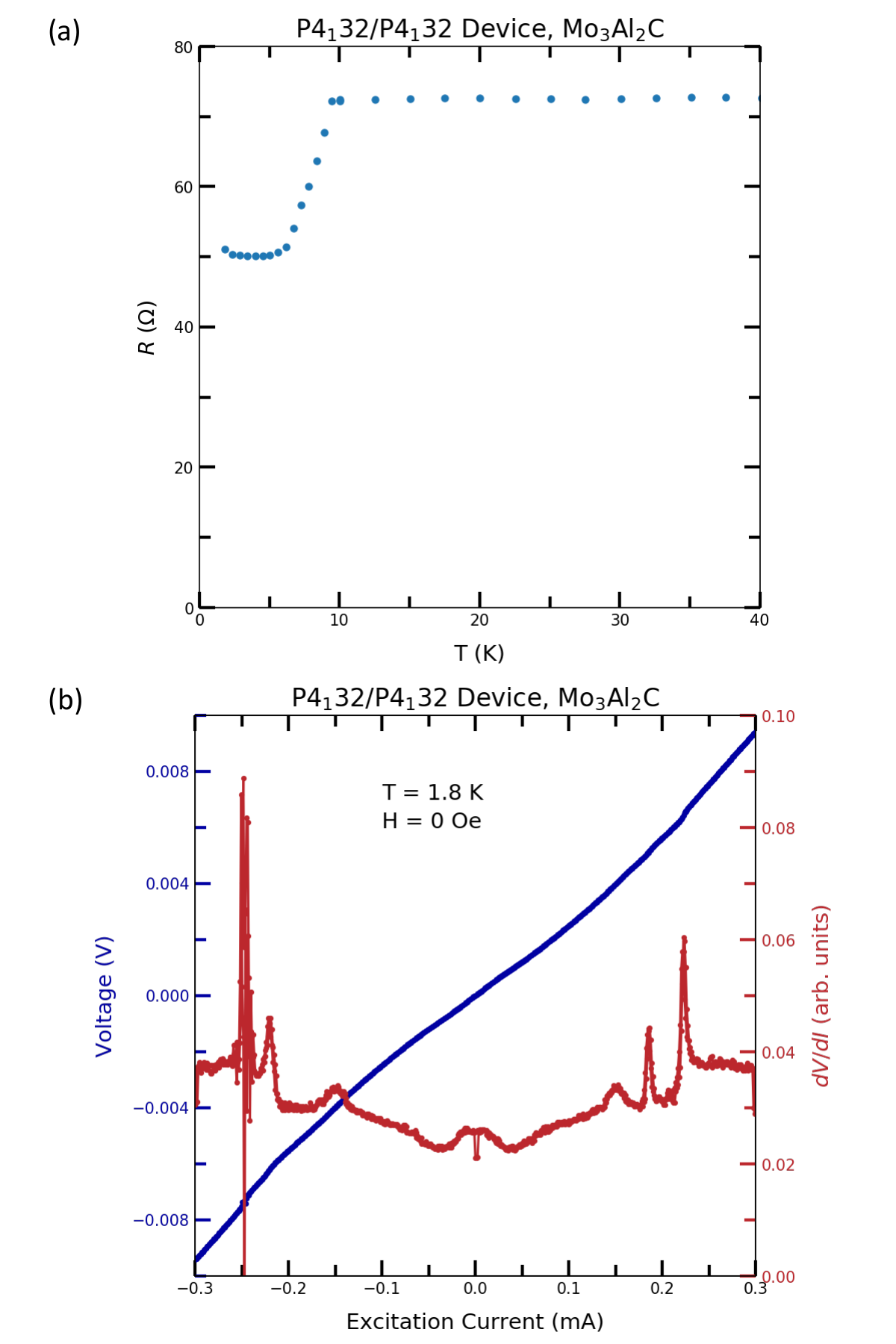}
\caption{(a) Resistance versus temperature and (b) voltage and $dV/dI$ versus excitation current for device 8.}
\end{figure}

\begin{figure}[H]
\centering
\includegraphics[width= 0.9 \linewidth]{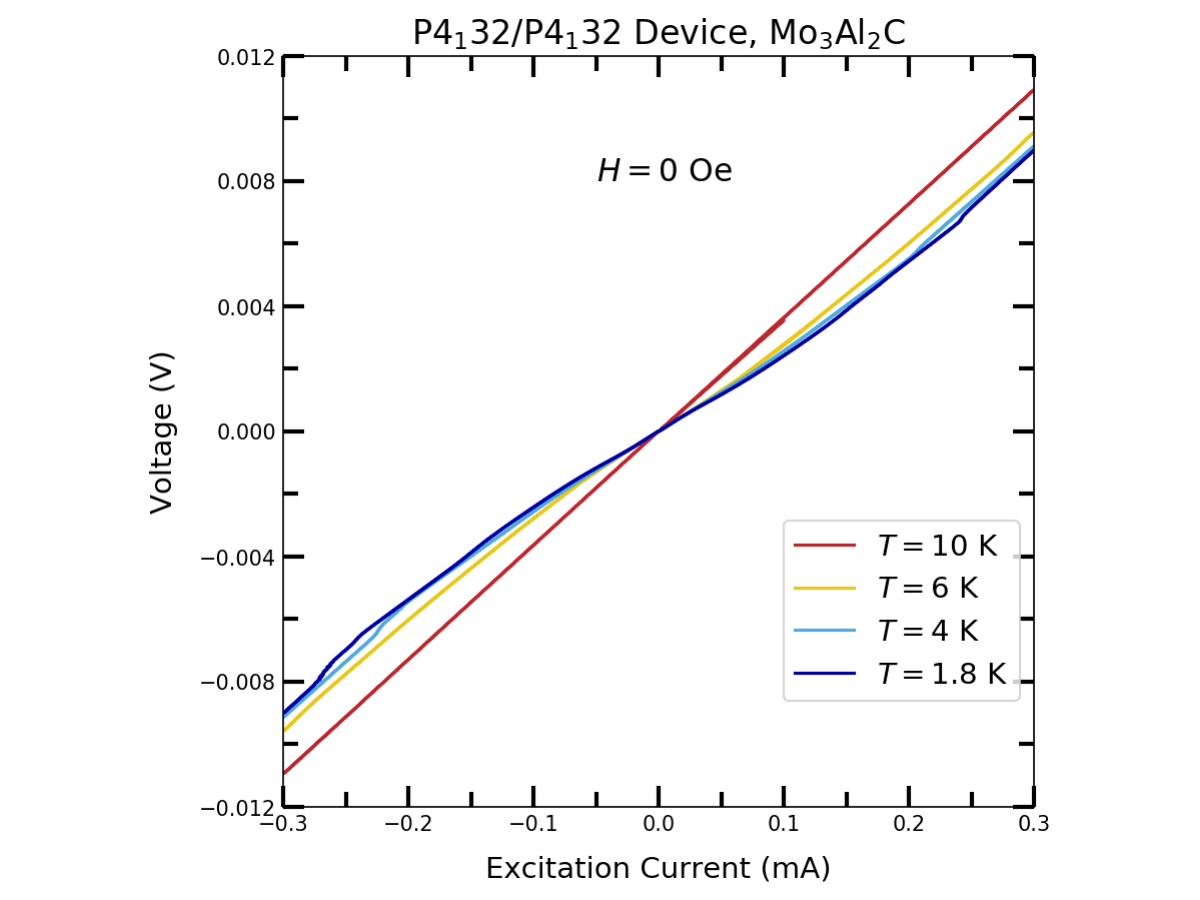}
\caption{IV curves at $T= 10, 6, 4,$ and $1.8$ K for device 8.}
\end{figure}

\section{Critical Current observed versus true critical current}
Due to contact resistance, we are not measuring the actual critical current of the Josephson junction. The observed transition is systematically shifted, as shown in SI Figure 20. The difference between $I_{c+}$ and $I_{c-}$ is not affected by the systematic shift, so measuring $I_{c+}^{*}$ and $I_{c-}^{*}$ can determine whether there is a diode effect. We define the critical current as the maximum value of $dV/dI$ in the IV curve.
\begin{figure}[H]
\centering
\includegraphics[width=\linewidth]{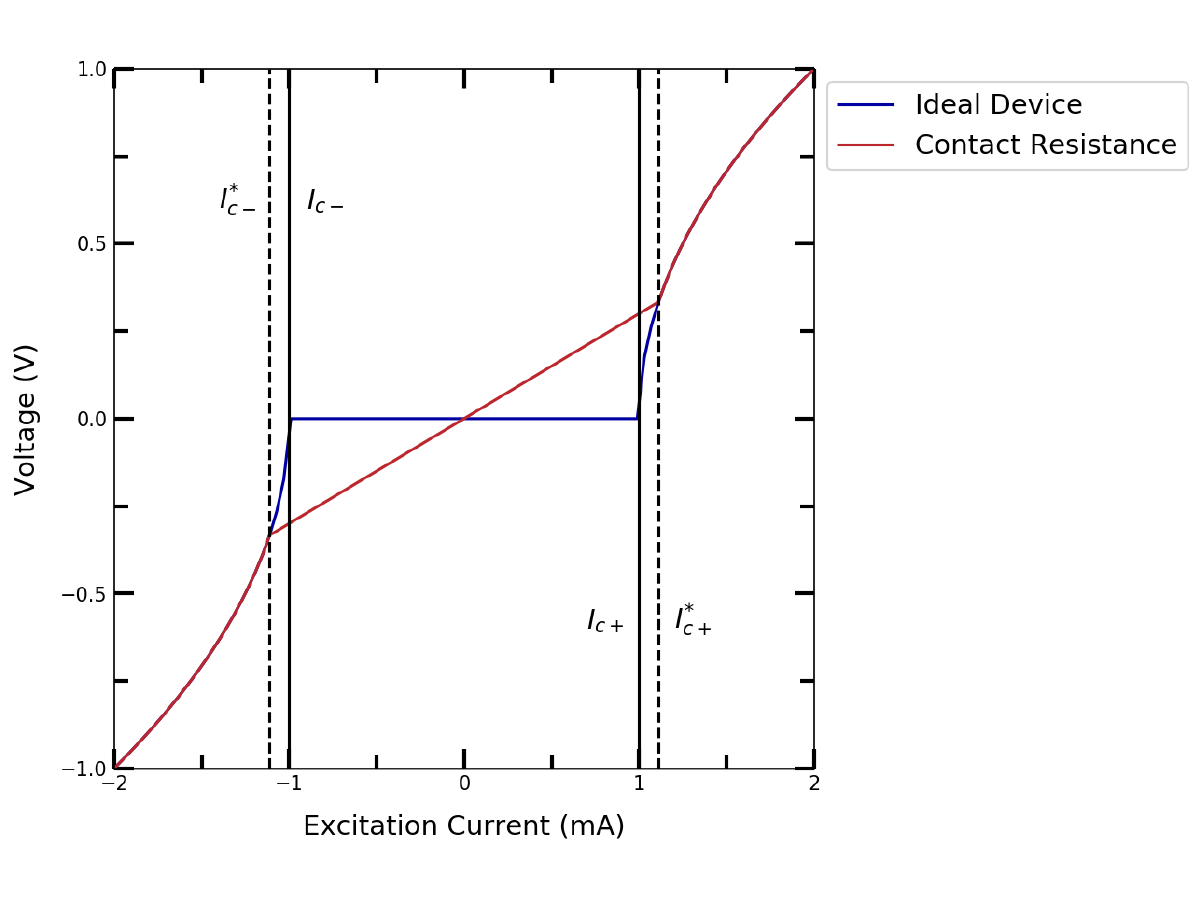}
\caption{IV curve for an ideal device and a device with contact resistance. Contact resistance makes it impossible to observe the true critical currents $I_{c \pm}$, however it is still possible to observe systematically shifted critical currents $I_{c \pm}^{*}.$}
\end{figure}

\section{Surface Characterization}
\subsection{Scanning Electron Microscopy}
\begin{figure}[H]
\centering
\includegraphics[width=0.8 \linewidth]{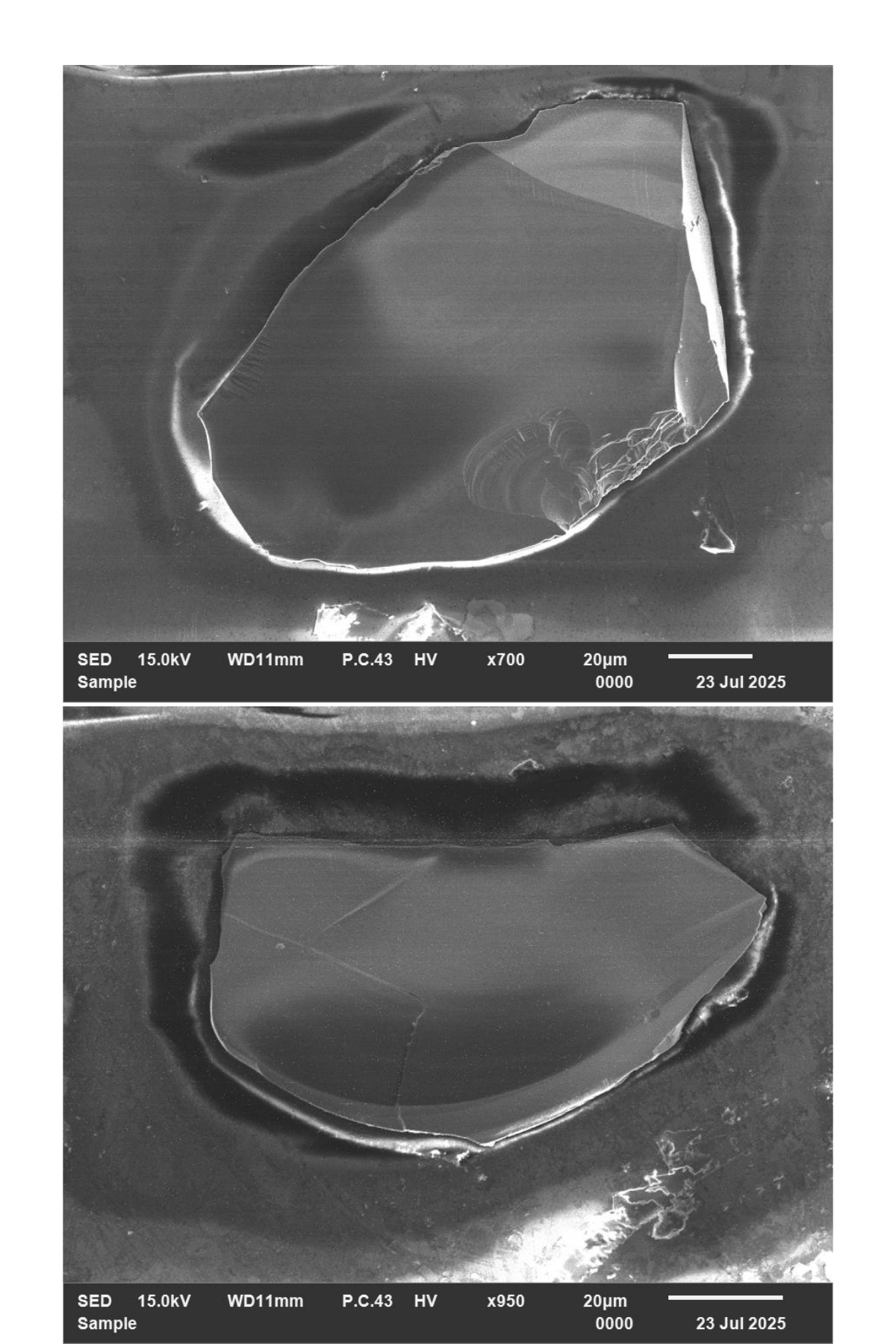}
\caption{SEM images of two single crystals of Mo$_3$Al$_2$C.}
\end{figure}

\begin{figure}[H]
\centering
\includegraphics[width= 0.9 \linewidth]{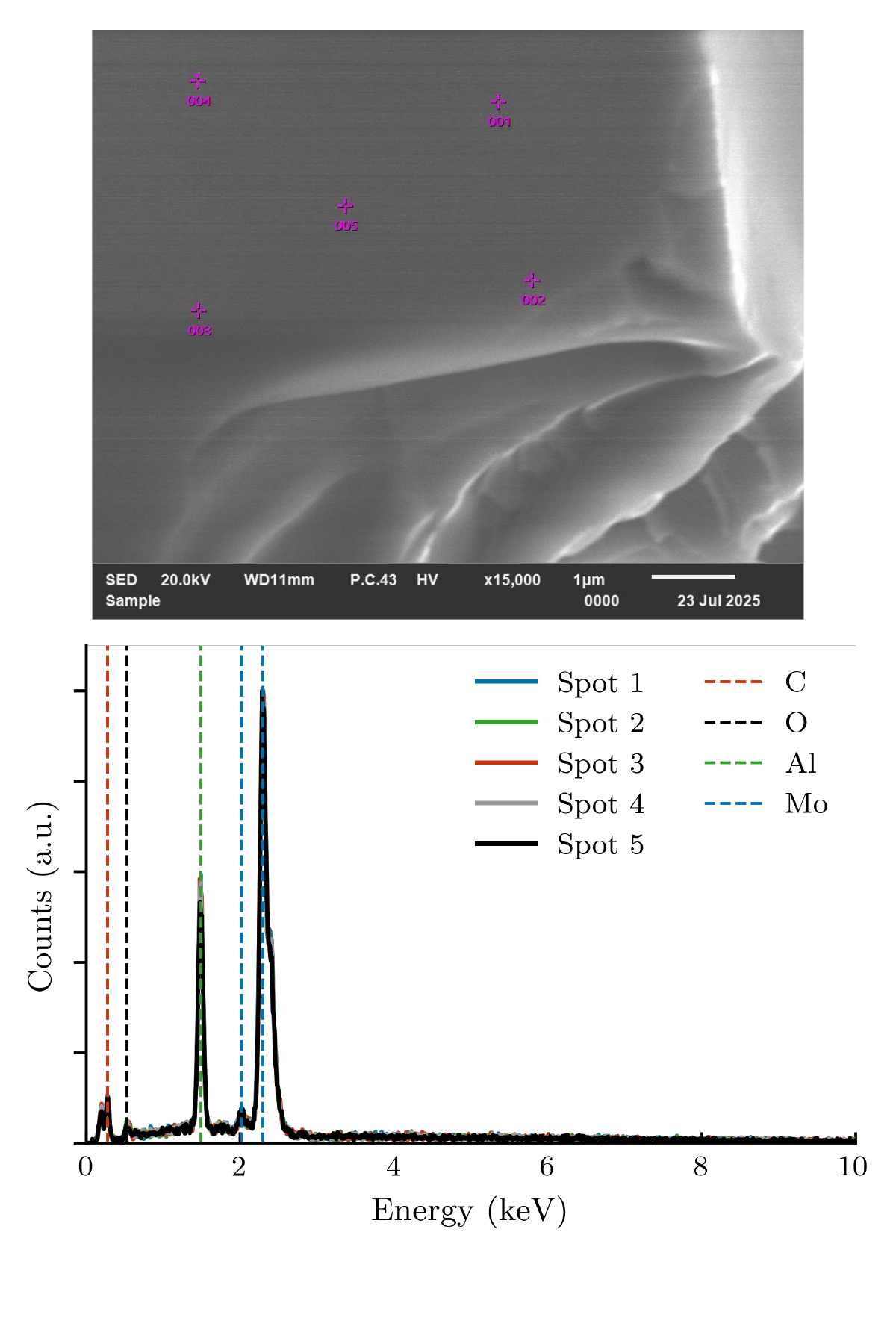}
\caption{Zoomed in SEM image of a Mo$_3$Al$_2$C crystal and EDS spectra.}
\end{figure}

\begin{table}[h]
\caption{Quantitative EDS spectra determining the atomic composition  of Mo, Al, C, and O at different spots on the Mo$_3$Al$_2$C single crystal}\label{tab1}%
\begin{tabular}{@{}lllll@{}}
\toprule
Mo$_3$Al$_2$C & Mo at \%  & Al at \% & C at \% & O at \% \\
\midrule
Spot 1    & 31.21   & 18.22  & 49.30 & 1.27  \\
Spot 2    & 33.99   & 19.38  & 37.57 & 9.06  \\
Spot 3   & 29.70   & 17.71  & 48.52 & 4.06  \\
Spot 4    & 28.80   & 17.19  & 48.36 & 5.65  \\
Spot 5    & 29.84   & 17.08  & 47.10 & 5.97  \\
Average    & 31(2)   & 18(1)  & 46(5) & 5(3)  \\\botrule
\end{tabular}

\end{table}






\end{document}